\documentclass[twocolumn,superscriptaddress,amsmath,amssymb,aps,prl]{revtex4-1}

\usepackage{graphicx}% Include figure files
\usepackage{bm}% bold math
\usepackage{natbib}
\usepackage{color}
\usepackage[colorlinks=true,bookmarks=false,citecolor=blue,urlcolor=blue]{hyperref} 

\newcommand{\ket}{\rangle}
\newcommand{\bra}{\langle}

\newcommand{\+}{\dagger}

\newcommand{\dd}{\mathrm{d}}

\newcommand{\re}{\mathfrak{Re}}

\newcommand{\op}{\widehat}

\newcommand{\hsig}{\op{\sigma}}

\begin{document}

\title{Strong coupling out of the blue: an interplay of quantum emitter hybridization with plasmonic dark and bright modes}

\author{Benjamin Rousseaux}
\email{benjamin.rousseaux@chalmers.se}
\affiliation{Department of Physics, Chalmers University of Technology, 412 96 G\"oteborg, Sweden}
\affiliation{Department of Microtechnology and Nanoscience - MC2, Chalmers University of Technology, 412 96 G\"oteborg, Sweden}

\author{Denis G. Baranov}
\affiliation{Department of Physics, Chalmers University of Technology, 412 96 G\"oteborg, Sweden}

\author{Tomasz J. Antosiewicz}
\affiliation{Faculty of Physics, University of Warsaw, Pasteura 5, 02-093 Warsaw, Poland}
\affiliation{Department of Physics, Chalmers University of Technology, 412 96 G\"oteborg, Sweden}

\author{Timur Shegai}
\affiliation{Department of Physics, Chalmers University of Technology, 412 96 G\"oteborg, Sweden}

\author{G\"oran Johansson}
\affiliation{Department of Microtechnology and Nanoscience - MC2, Chalmers University of Technology, 412 96 G\"oteborg, Sweden}

\begin{abstract}
Strong coupling between a single quantum emitter and an electromagnetic mode is one of the key effects in quantum optics. In the cavity QED approach to plasmonics, strongly coupled systems are usually understood as single-transition emitters resonantly coupled to a single radiative plasmonic mode. However, plasmonic cavities also support non-radiative (or ``dark'') modes, which offer much higher coupling strengths. On the other hand, realistic quantum emitters often support multiple electronic transitions of various symmetry, which could overlap with higher order plasmonic transitions -- in the blue or ultraviolet part of the spectrum. Here, we show that vacuum Rabi splitting with a single emitter can be achieved by leveraging dark modes of a plasmonic nanocavity. Specifically, we show that a significantly detuned electronic transition can be hybridized with a dark plasmon pseudomode, %which is strongly detuned from the bright dipolar mode,
resulting in the vacuum Rabi splitting of the bright dipolar plasmon mode. We develop a simple model illustrating the modification of the system response in the ``dark'' strong coupling regime and demonstrate single photon non-linearity. These results may find important implications in the emerging field of room temperature quantum plasmonics. 
%\red{the abstract gives a wrong impression. TS: yeah, the point is that the splitting is given by g-bright anyhow, so we simply use the fact that high dipole moment molecules are typically in the blue. We utilize the dark mode to shift its high transition dipole resonance to the visible range due to coupling to the dark mode, where it can overlap with the bright mode.}
\end{abstract}
\maketitle

\textit{Introduction}.---
Interaction of a quantum emitter (QE) with an optical cavity is at the heart of modern quantum optics. In the regime of weak QE-cavity coupling the presence of a QE may be treated as a perturbation that affects the eigenmode of the cavity \cite{Scully,Tame}. However, when the interaction between the cavity mode and the QE is strong enough, they form dressed polaritonic states separated by the vacuum Rabi splitting in the energy spectrum \cite{Khitrova2006, Fink2008, Torma2015, baranov2017novel}.
As the QE and the optical mode can no longer be treated as separate entities in this regime, such an evolution of the system not only modifies its optical response, but also dramatically affects exciton transport~\cite{Transport} and photochemical \cite{Angew2012, Angew2016, Chemistry2016, Galego2016, munkhbat2018suppression} properties.

Strong light-matter coupling is particularly interesting in the single emitter limit, when unique features of the Jaynes-Cummings ladder enable single-photon optical nonlinearities \cite{Birnbaum2005,Englund2007}. Rabi splitting between single quantum dots and dielectric high-Q microcavities was observed in a number of works, but only at cryogenic temperatures \cite{Yoshie2004,Reithmaier2004}. Plasmonic nanocavities enable observation of strong coupling with quantum dots and organic chromophores at room temperatures \cite{Chikkaraddy2016,Santhosh2016,gross2018near,PeltonNat18}, but most of such structures are at the border between the weak and the strong coupling regime due to limited coupling strength \cite{rousseaux2018comparative}. 

The value of the coupling strength is determined by the transition dipole moment of the QE and the vacuum electric field of the cavity \cite{Khitrova2006, Torma2015, baranov2017novel}. To achieve Rabi splitting in the visible range, the electronic transition of the QE has to be resonant with the bright mode of the cavity in the visible range. However, many material systems that are used to emulate QEs, for example colloidal quantum dots \cite{leatherdale2002absorption, QDreview2003} and excitons in transition metal dichalcogenides monolayers \cite{wang2017excitons}, also possess electronic transitions at higher energies, which are often characterized by higher values of the oscillator strength.
The high oscillator strength of these transitions could potentially be used to enhance the magnitude of Rabi splitting if the dipolar plasmon resonance can be tuned to the appropriate frequency range to overlap with those transitions. However, such approach would require tuning dipolar plasmon resonances to the UV range, which has a number of disadvantages, including the complexity of optical measurements in this spectral range and the necessity of utilizing metals with significantly high plasma frequency, such as aluminium \cite{knight2013aluminum, rossi2019}. 

Alternatively, one could explore the possibility of strong coupling between the QE and the so-called ``dark'', non-radiative modes of conventional Ag and Au nanoparticles \cite{liu2009excitation,Delga_2014,RousGuer16,VargCola16,LiFern18,CastGuer18,VargCola19,CuarFernPre}. Despite the fact these dark modes are not observable using traditional optical techniques (although can be observed by EELS \cite{koh2009electron, barrow2014mapping,BittHaraPre}), it might be possible to visualize them by further hybridization of the dark mode-QE state with the bright mode of the resonator.
In the weak coupling scenario, the interaction of a QE with a dark mode leads to quenching of emission \cite{LukasNovotny2012,Anger2006}, which is why these modes are often assumed to be detrimental for the purposes of vacuum Rabi splitting. In the strong coupling regime, however, Rabi splitting is relatively robust with respect to quenching when the emitter is spectrally tuned to the bright dipole mode of a plasmonic nanoparticle as was shown recently \cite{Delga_2014}. It has also been shown that light-forbidden quadrupolar transitions of excitons coupled to a nanoparticle on mirror system can lead to strong coupling \cite{CuarFern18,CuarFern19}.

\begin{figure}[b]
\includegraphics[width=1\columnwidth]{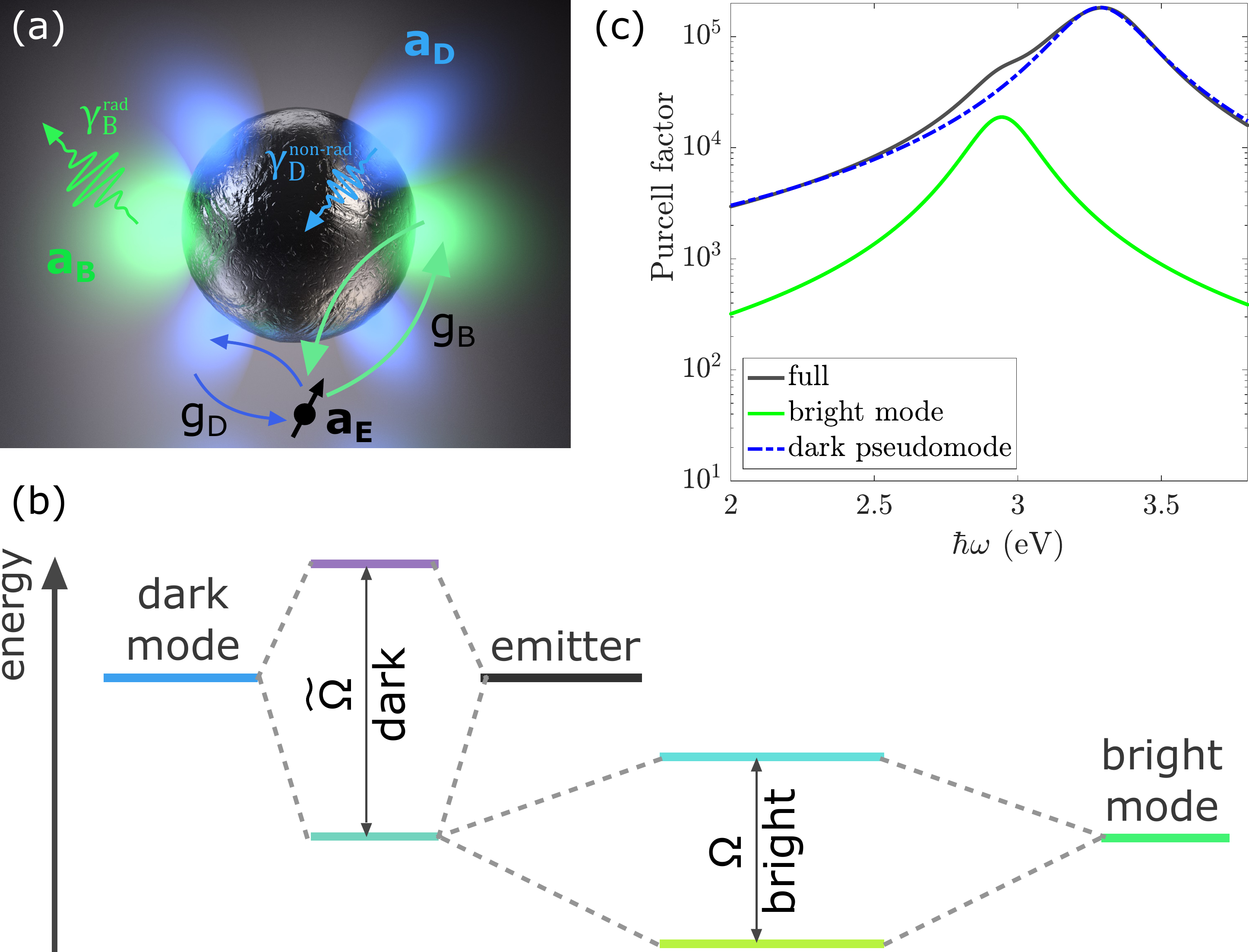}
\caption{(a) Schematic illustration of the system under study: a cavity with a bright mode and a dark mode couples to a QE. (b) Sketch of the energy diagram of the three-component system. The dark mode-emitter coupling results in ``dark'' polaritons; the lower of them two in turn couples with the bright mode, resulting in two ``bright'' polaritons that can be resolved in the scattering spectrum. (c) LDOS spectrum for a 10 nm Ag nanosphere, 1 nm away from the surface. The green line shows the dipole mode contribution, while the dashed line is the dark pseudomode contribution.}
\label{fig1}
\end{figure}

\begin{figure*}
\includegraphics[width=1.9\columnwidth]{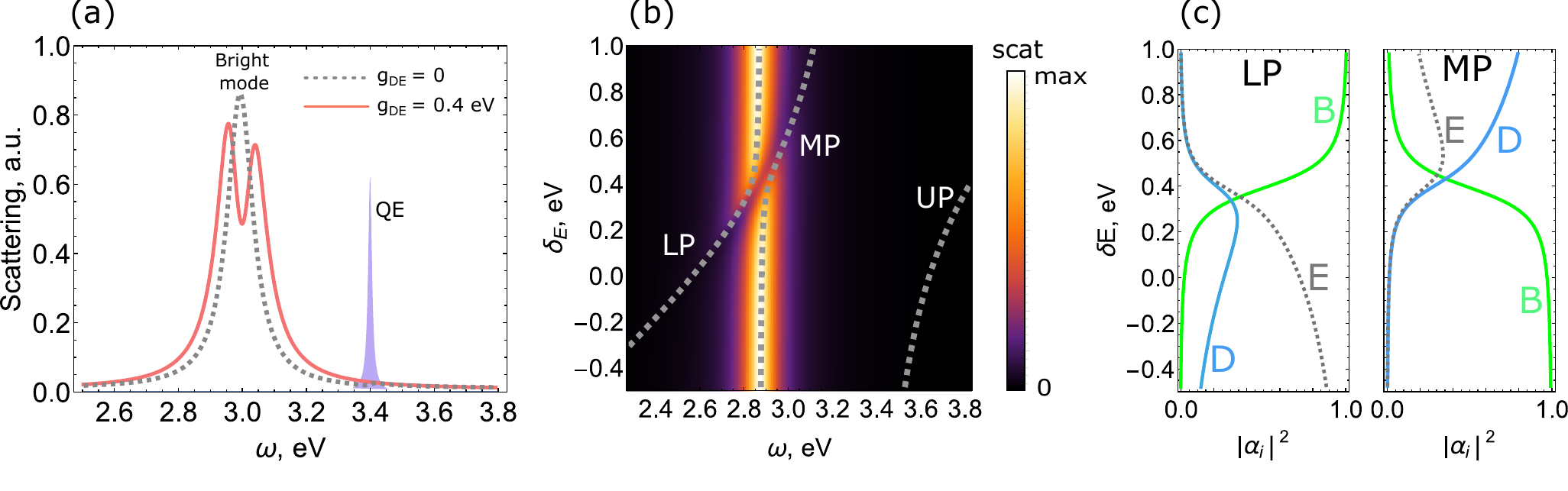}
\caption{Theoretical modelling of a generic coupled three-component system. (a) Scattering spectrum by the 3$\times$3 system with the exemplary parameters outlined in the text in the absence of the dark mode, $g_{DE}=0$ (dashed), and in the presence of the dark mode-emitter coupling, $g_{DE}=0.4$ eV (solid). The filled curve depicts spectral position of the QE transition at 3.4 eV. (b) Map of simulated scattering spectra $|s_-|^2$ versus emitter detuning $\delta_E$ for the exemplary 3$\times$3 system. The dashed lines show the real parts of the system eigenfrequencies. (c) Hopfield coefficients of the lower (left) and middle (right) polaritons versus emitter detuning. 
}
\label{fig2}
\end{figure*}

In this Letter, we demonstrate theoretically that by coupling a high-energy transition of a QE to a cavity dark mode, it is possible to achieve observable Rabi splitting between two bright polariton modes. The dark mode plays a role of a tuning mechanism of the high-energy QE resonance towards the bright plasmon mode, where the interaction can take place.
We analyze the system response with the use of a generic coupled mode system, as well as Green's tensor calculations for spherical geometry and a master equation approach.  Noteworthy, all the parameter values correspond to a realistic geometry, thereby suggesting a practical recipe for the realization of vacuum Rabi splitting with a single QE at room temperature. Our results could potentially help in understanding the microscopic behaviour of experimental observations of QE-plasmon systems such as shown in refs. \cite{Santhosh2016,gross2018near}.

%\red{(TS: By the way, Hecht is using Gold! So our theory might not explain this, since in Gold all dark modes are suppressed due to inter-band transitions. So, not sure how to treat his data. Only may be that this broadband antenna has some dark modes even where Gold still does not absorb much...)}

%We demonstrate \red{single photon non-linearity}, and suggest an experimentally feasible structure for realization of this effect (\red{DGB: nanosphere? I doubt it's feasible. Other than that, we don't suggest any structures}).

\textit{Results}.---
The system under study is schematically shown in Fig. \ref{fig1}(a). It is composed of a generic optical cavity and a QE. The cavity has two modes, one of which is bright (``B'') and has low non-radiative loss $\gamma_B^\text{non-rad}$, while the other one is dark (``D'') and has low radiative loss $\gamma_D^\text{rad}$. The emitter couples to the bright mode and to the dark pseudomode with coupling strengths $g_B$ and $g_D$, respectively. The energy diagram sketched in Fig. \ref{fig1}(b) elucidates the resulting interaction picture in this kind of system. The emitter interacts with the dark mode resulting in two polariton modes separated by a ``dark'' Rabi splitting, which can not be observed in the far field. The lower of these two polaritons, in turn, interacts with the bright cavity mode leading to formation of another pair of polaritonic states, which can be observed in scattering owing to the radiative character of the bright mode.

First, we apply a simple analytical model based on the temporal coupled mode theory to our system  \cite{Haus,Fan2003}. This model captures the most important features of the system response. In this framework, the system response is described by a ket-vector with complex amplitudes $\left| a \right\rangle  = {\left( {{c_B},{c_D},{c_E}} \right)^T}$, where the subscripts $B,D,E$ denote corresponding amplitudes for the bright mode, the dark pseudomode and the QE, respectively.
The dynamics of the amplitudes is governed by the Schr\"odinger-like equation
\begin{align}
i\frac{{d\left| a \right\rangle }}{{dt}} = \widehat H\left| a \right\rangle  +{s_ + } \left| \kappa  \right\rangle ,\quad \left| \kappa  \right\rangle = {\big( \kappa_B ,\kappa_D ,\kappa_E \big)^T}
\label{eq1}
\end{align}
where $\widehat H$ is the system Hamiltonian, $\left| \kappa  \right\rangle$ is the mode-radiation coupling constants vector with components $\kappa_j = \sqrt{\gamma_j^\text{rad}}$, $\gamma_j^\text{rad}$ are the radiative decay rates of each mode, and ${s_ + }$ is the incident wave amplitude. The Hamiltonian of the three mode system reads:
%%%
\begin{equation}
\widehat H = \left( {\begin{array}{*{20}{c}}
{{\omega _B} - i{\gamma _B/2}}&0&{g_B - i{\gamma _{{\rm{ind}}}}}\\
0&{{\omega _D} - i{\gamma _D/2}}&{{g_D}}\\
{{g_B} - i{\gamma _{{\rm{ind}}}}}&{{g_D}}&{{\omega _E} - i{\gamma _E/2}}
\end{array}} \right),
\label{eq2}
\end{equation}
%%%
where $\omega_j,\gamma_j$ stand for the eigenfrequencies and total decay rates of each mode, respectively. The non-Hermitian term with $\gamma _{\rm{ind}} =\sqrt {\gamma _B^{{\rm{rad}}}\gamma _E^{{\rm{rad}}}/4}$ comes from the far-field (indirect) coupling of the bright mode with the QE \cite{suh2004temporal} and can be neglected when the QE radiative decay is much smaller than that of the bright mode. For a harmonic excitation at frequency $\omega$, the steady state solution of Eq. 1 reads $\left| a \right\rangle  = \frac{{\left| \kappa  \right\rangle {s_ + }}}{{i(\widehat H - \omega )}}$.
Finally, the amplitude of the scattered signal in the steady state regime is given by ${s_ - } = \left\langle \kappa | a \right\rangle $.
%For the inter-mode coupling coefficients, modifying normal modes of Hamiltonian (2), we will take $g_{{\rm{BE}}=0.1 \omega_B$, whereas $g_{{\rm{DE}}$ will be varied.
We consider a cavity with the bright mode at 3 eV, and the dark pseudomode at 3.4 eV, corresponding to an Ag nanosphere of 10 nm diameter, Fig. \ref{fig1}(c). For the bright and dark mode linewidths we will use $\gamma_B^{\rm{rad}}=\gamma_D^{\rm{non-rad}}=0.05$ eV. Furthermore, we will assume $\gamma _D^{{\rm{rad}}} = \gamma _B^{{\rm{non-rad}}} =0$.
To strengthen our motivation, we examine which QEs might be suitable for the proposed strong coupling scheme. Colloidal quantum dots (QDs), such as CdSe QDs, have a transition dipole moment of about 5-15 D at the wavelength of 600 nm \cite{leistikow2009size}. At the same time, these QDs are known to have high absorption and extinction coefficients in the UV range, exceeding that in the visible range by at least an order of magnitude \cite{leatherdale2002absorption,yu2003experimental}. Recalling that the extinction cross-section of a two-level system is related to its transition dipole moment $\mu$ via ${\sigma _{ext}} = \left( {{\omega _E}{\mu ^2}} \right)/\left( {\hbar c{\varepsilon _0}{\gamma _E}} \right)$ \cite{leatherdale2002absorption}, where $c$ is the speed of light, $\varepsilon_0$ the vacuum permittivity, and assuming that the absorption peak predominantly originates from a single electronic transition (which might be not true in a realistic system), we may realistically estimate the dipole moment of the UV transition is of the order of 100 D. Based on this simple estimation, we assign $g_B=0.05$ eV and $g_D=0.4$ eV, corresponding to a point emitter located 1 nm from the surface of the Ag nanosphere. According to the Larmor formula for the radiative decay rate, this value of the transition dipole moment results in $\gamma_E^{\rm{rad}} \approx 3$ $\mu$eV, what is negligible in comparison to other decay rates.

To gain initial understanding of the three-component system behavior, we examine in Fig. \ref{fig2}(a) how the presence of the dark mode affects the elastic scattering spectrum for the QE tuned to the dark mode energy of 3.4 eV in accordance with Eqs. (1-2). When the dark mode is turned off, $g_D=0$, the scattering spectrum exhibits one prominent peak corresponding to the uncoupled bright mode. However, when the coupling to the dark mode is introduced via $g_D$, the scattering spectrum presents two peaks around 3 eV suggesting the onset of strong coupling between the emitter and the bright mode.

In order to corroborate the strong coupling regime upon coupling to the dark mode, we analyze the elastic scattering from the system versus the QE detuning $\delta_E=\omega_E-\omega_B$, Figs. \ref{fig2}(b).
As one can see, an anti-crossing occurs when the QE frequency crosses the dark mode frequency ($\delta_E \approx 0.4$ eV) i.e. at $\omega_E \approx \omega_D$. Notably, the Rabi splitting itself still occurs at the frequency of the unperturbed bright mode around 3 eV. The scattering peaks precisely follow eigenenergies of the Hamiltonian (Eq. \ref{eq2}), which are shown by the dashed lines in Fig. \ref{fig2}(b). The anti-crossing of the eigenvalues confirms the strong coupling regime in the system.
This is the main result of our letter that we would like to emphasize: one can leverage high transition dipole moments of certain QEs typically lying in the UV region for enhanced Rabi splitting in the visible range, provided that the emitter additionally interacts with a high-energy non-radiative mode.

%To better understand these plots, one should keep in mind that, although the scattering map with $g_{\rm{DE}}=0$ also demonstrates vacuum Rabi splitting around 2 eV, this splitting occurs for a QE with a low energy transition ($\delta_e=0$), whose strength is typically much weaker that that of higher energy transitions (\red{In other words, it corresponds to a not so realistic situation}). 

\begin{figure*}
\includegraphics[width=2.1\columnwidth]{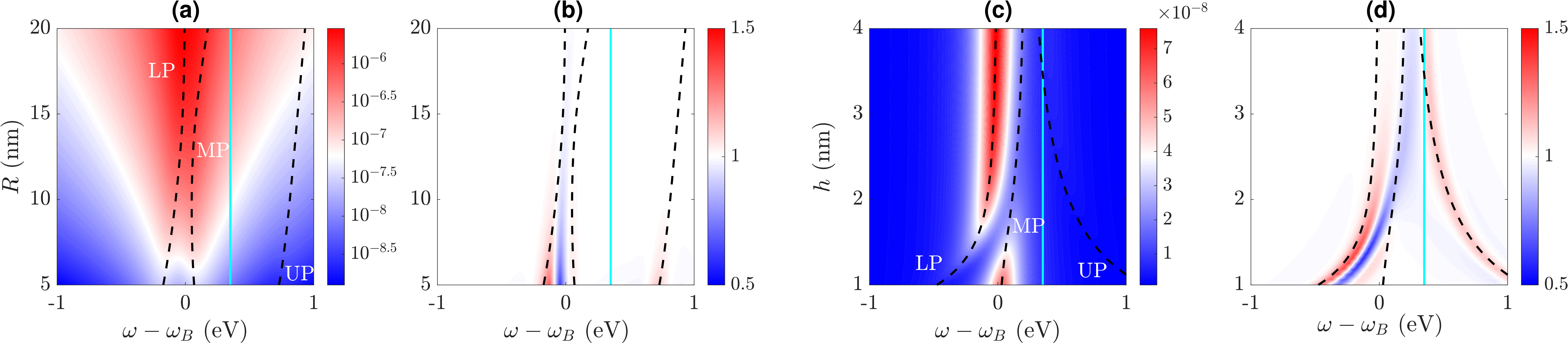}
\caption{Strong dark coupling in an Ag nanosphere. (a) False-color map of calculated scattering spectra of a 100 D point dipole QE 1.5 nm away from the surface of an Ag nanosphere of radius $R$. The emitter frequency is set to $\omega_E = \omega_D$, 0.35 eV away from the bright mode (cyan vertical line). For better visibility, this map is shown in log scale. (b) Second order correlation function $g^{(2)}(0)$ versus $R$ for $h=$ 1.5 nm. (c) Calculated scattering spectra for the same system versus the emitter-surface distance $h$ for $R=$ 5 nm. (d) Second order correlation function $g^{(2)}(0)$ versus $h$ for $R=$ 5 nm.}
\label{fig3}
\end{figure*}

Eigenvectors of Hamiltonian (Eq. \eqref{eq2}) correspond to three-component quasiparticles: $\left| {E{P_i}} \right\rangle  = {c_B}\left| B \right\rangle  + {c_D}\left| D \right\rangle  + {c_{E}}\left| {E} \right\rangle$, where $\left| B \right\rangle$, $\left| D \right\rangle$, and $\left| E \right\rangle$ denote the bare bright mode, dark mode, and QE states, respectively. The lowest, medium, and highest energy solutions are referred to as the lower, middle, and upper polaritons (LP, MP, UP), respectively. Absolute amplitudes of these contributions (Hopfield coefficients), shown in Fig. \ref{fig2}(c) for the LP and MP as a function of the QE detuning, confirm that both bright polaritons have contributions from the bright and dark mode as well as the QE at the avoided crossing position and thus indeed present mixed light-matter states.
The ``bright'' Rabi splitting observed in the spectra around 3 eV occurs between the LP and MP. Neglecting losses, we can obtain an analytical expression for the magnitude of this splitting from the Hamiltonian (Eq. \ref{eq2}) (see Supporting Information):
%%%
\begin{equation}
\Omega_{\rm bright} = \sqrt{2}g_B\sqrt{1 + \frac{\omega_D - \omega_E}{\sqrt{4g_D^2 + (\omega_D - \omega_E)^2}}}.
\end{equation}
%%%
As one can see, it is mostly affected by the bright-emitter coupling constant $g_B$, which is determined by the transition dipole moment of the emitter and the vacuum electric field of the lower-energy bright mode \cite{Khitrova2006,baranov2017novel}. The dark-emitter coupling constant $g_D$, at the same time, has a negative effect on the resulting splitting. However, it is the large coupling to the dark mode that allows to effectively "push" the QE resonance down to the visible region, where it can interact with the bright mode. This role of the dark mode-emitter coupling can be illustrated by the expression for the optimal emitter-bright mode detuning $\delta _E^{\rm{opt} }$, upon which the bright mode is in zero detuning with the polariton formed by the dark mode-QE coupling (see Supporting Information for details):
%%%
\begin{equation}
\delta _E^{\rm{opt} } = g_D^2/\left( {{\omega _D} - {\omega _B}} \right)
\end{equation}
%%%
Essentially, this equation shows that the larger the dark mode-emitter coupling is, the higher $\omega _E$ should be in order for its hybridized resonance to overlap perfectly with the bright mode in the visible region.

We further elaborate the concept of dark strong coupling by inspecting the response of a specific nanocavity, with the use of an effective master equation approach (see Supporting Information for details). We choose a silver spherical nanoparticle of radius $R$ and a QE placed at a distance $h$ from the nanosphere surface. As was mentioned above, the dipole moment of the UV transition of some QEs could reach 100 D ($\sim$2 e$\cdot$nm), and the total decay rate of such a transition to be of the order of 0.1 eV. The map of elastic scattering versus the nanoparticle radius presented in Fig. \ref{fig3}(a) confirms that the Rabi splitting due to the dark mode coupling is preserved for a wide range of the nanoparticle size. In this plot, the QE detuning was placed in resonance with the dark pseudomode so that $\delta _E = 0.35$ eV, for the smallest radius of 5 nm. 
The observed effect appears to be much more sensitive to the surface-emitter separation $h$, as Fig. \ref{fig3}(b) indicates. The Rabi splitting in the vicinity of the bright mode is sustained only up to 2 nm separation and disappears for larger distances, where only the uncoupled bright mode and the emitter contribute to scattering. This behavior originates from the dark pseudomode strong dependence on $h$. With increasing $h$, the coupling to the dark mode quickly diminishes, leaving only the signatures of the bright mode in the spectrum.

%In Fig. 3(c) we inspect the evolution of the scattering spectrum while scanning the emitter detuning with respect to the bright mode $\delta_E$. The resulting picture exhibits a clear anti-crossing near the optimal detuning in a manner very similar to Fig. 2(b), what confirms \red{XXX} and justifies the use of the simple 3$\times$3 model.

Finally, we demonstrate the photon blockade for $R = 5$ nm versus $h$. The results are shown in Fig. \ref{fig3}(c), where we plot the scattered photon statistics for zero delay, i.e. the second order correlation function \cite{SaezGarc17}:
\begin{align}
\label{g2}
g_{\omega}^{(2)}(0) = \frac{\bra\op{a}_s^\+\op{a}_s^\+\op{a}_s\op{a}_s\ket}{\bra\op{a}_s^\+\op{a}_s\ket^2},
\end{align}
where $\op{a}_s = \mu_E\hsig_- + \mu_B\op{a}_B$ is the scattered light operator, $\hsig_-$ is the lowering operator of the QE transition, $\mu_B$ is the nanosphere dipole moment and $\op{a}_B$ is the bright mode annihilation operator. We show that antibunched light ($g_\omega^{(2)}(0) < 1$) is produced following the LP since it is a mixture of the dark plexciton and the bright mode, while slightly bunched light ($g_\omega^{(2)}(0) > 1$) appear on the dark plexciton UP. We underline here that the antibunching is resulting from strong interactions with both dark and bright modes, even if the dark pseudomode is usually thought of as being detrimental for the radiative properties of the system. Also, even if the QE is being hybridized with two plasmon modes, the photon statistics shows clear antibunching, indicating the robustness of single photon emission in this scheme. Finally, despite the resonances being in the near UV, the single photon emission line is shown to be red-shifted so that it can be seen in the visible. We further discuss this effect in the Supporting Information.

\textit{Conclusion}.---
We have presented a novel scheme for realizing strong light-matter coupling with use of a high-energy electronic transition of a large oscillator strength quantum emitter. Exploiting the non-radiative modes of a plasmonic cavity, the high-energy transition can be tuned to lower energies, where it can couple with the bright plasmon cavity mode leading to observable vacuum Rabi splitting in the scattering spectrum. Results were predicted by a simple model and verified with the use of an effective master equation approach for realistic coupling parameters and cavity geometries. Quantum nonlinearities were also shown with the use of the second order coherence function and found to be robust with respect to dark mode coupling. UV transitions of colloidal quantum dots or C-excitons of transition metal dichalcogenides are possible candidates for the proposed approach towards strong coupling \cite{LeatBawe02,LiHein14}. This work could help in the design of novel QE-plasmon coupling schemes towards the realization of efficient room temperature strong coupling and quantum nonlinearities.

\begin{acknowledgments}
The authors acknowledge support from the Swedish Research Council (VR grant number: 2016-06059).
\end{acknowledgments}

\newpage
\onecolumngrid
\appendix
\section*{SUPPORTING INFORMATION}
We provide supporting information about the calculation of the scattering maps as well as details on formulas (3) and (4) using a partial diagonalization approach for the Hamiltonian. The latter is constructed using a mode decomposition for the plasmonic resonances of the spherical nanoparticle. The construction of this model is well understood in the framework of the Green's tensor approach \cite{HakaZuba14,RousGuer16,DzsoGuer16}. 

  \section{Bright and dark mode decomposition - effective Hamiltonian}
In the rotating wave approximation, the non-Hermitian Hamiltonian for the nanosphere-emitter system reads, using the spherical orthogonal mode decomposition:
\begin{align}
\label{Hnh}
\widetilde{H}_\text{m.d.} = \Big(\omega_E - i\frac{\gamma_E}{2}\Big)\hsig_+\hsig_- + \sum_{n = 1}^{\infty}\Big(\omega_n - i\frac{\gamma_n}{2}\Big)\op{a}_n^\+\op{a}_n + \sum_{n=1}^{\infty}g_n(\op{a}_n^\+\hsig_- + \op{a}_n\hsig_+),
\end{align}
where $\omega_E$ is the transition frequency of the QE, $\hsig_-,\hsig_+$ its lowering and raising operators, respectively, and $\gamma_E$ its total decay rate. The plasmonic field is modeled with creation and annihilation operators $\op{a}_n^\+,\op{a}_n$ associated with frequencies $\omega_n$ and decay rates $\gamma_n$. Each $n$ mode corresponds to a specific plasmon resonance: $n=1$ is the dipolar mode, $n=2$ the quadrupolar, $n=3$ the octupolar and so on. In the case of a spherical nanoparticle, the dipole mode is usually well separated from the higher order modes $n\geqslant 2$ and the latter being quasi-degenerate behave effectively as a large pseudomode when the emitter is very close to the surface of the sphere. In the following we note $\omega_1\equiv \omega_B$, $\gamma_1 \equiv \gamma_B$, $\op{a}_1 \equiv \op{a}_B$, $g_1 \equiv g_B$ and:
\begin{align}
\op{a}_D = \frac{1}{g_D}\sum_{n\geqslant 2}^\infty g_n\op{a}_n.
\end{align}
The commutation relation of the original modes $[\op{a}_n,\op{a}_m^\+]= \delta_{nm}$ leads to the effective dark coupling to be $g_D = \left(\sum_{n\geqslant 2}^\infty g_n^2\right)^{1/2}$ in order to have the dark modes normalized and the right commutation relation $[\op{a}_D,\op{a}_D^\+]= 1$. The effective non-Hermitian system Hamiltonian then has the form:
\begin{align}
\label{Hnheff}
\widetilde{H}_S = \Big(\omega_E - i\frac{\gamma_E}{2}\Big)\hsig_+\hsig_- + \Big(\omega_B - i\frac{\gamma_B}{2}\Big)\op{a}_B^\+\op{a}_B + \Big(\omega_D - i\frac{\gamma_D}{2}\Big)\op{a}_D^\+\op{a}_D + g_B(\op{a}_B^\+\hsig_- + \op{a}_B\hsig_+) + g_D(\op{a}_D^\+\hsig_- + \op{a}_D\hsig_+).
\end{align}
The resonance $\omega_D$ and the decay rate $\gamma_D$ are obtained by fitting the pseudomode by a Lorentzian function and extracting its maximum position and full width at half maximum. The calculation of the Lorentzian-fitted LDOS from the Green's tensor approach then yields the parameters $(g_B,g_D,\omega_B,\omega_D,\gamma_B,\gamma_D)$ that appear in the non-Hermitian Hamiltonian.

\section{3$\times$3 Hamiltonian description: partial diagonalization and effective parameters}

The Hermitian part of the Hamiltonian \eqref{Hnh} can be written in a matrix form considering the single excitation basis: one excitation only is exchanged between the QE transition and the plasmon modes. Let the matrix form of the Hamiltonian generally be written in the basis $\{|e,0,0\ket,|g,1_D,0\ket,|g,0,1_B\ket\}$:
\begin{align}
\label{3x3}
\rm{H} = \left[\begin{array}{ccc} 0 & g_D & g_B\\g_D & \Delta_D & 0\\g_B & 0 & \Delta_B\end{array}\right],
\end{align}
where we wrote the Hamiltonian in a rotating frame with respect to $\omega_E$ , so that $\Delta_{D,B} = \omega_{D,B} - \omega_E$.
When an excitonic transition strongly couples to a plasmon mode, two polaritons (lower polariton (LP) and upper polariton (UP)) are formed and it is convenient to diagonalize the Hamiltonian block involving them. Also, writing the Hamiltonian in the basis of the polaritons enables to understand how the latter effectively couple to the other components of the Hamiltonian.

  \subsection{Diagonalization of the strongly coupled block}
In the following we consider the block $\Pi$ of the Hamiltonian \eqref{3x3}:
\begin{align}
\Pi = \left[\begin{array}{cc}0 & g_D\\g_D & \Delta_D\end{array}\right].
\end{align}
The eigenvalues of this block are the following:
\begin{subequations}
\begin{align}
\label{eigenv}
\delta_\pm &= \frac{1}{2}(\Delta_D \pm \Upsilon),\\
\Upsilon &= \sqrt{\Delta_D^2 + 4g_D^2}.
\end{align}
\end{subequations}
It is convenient to introduce the angle $\theta$ parametrized as following:
\begin{subequations}
\begin{align}
\cos\theta &= \frac{\Delta_D}{\Upsilon},\\
\sin\theta &= \frac{2g_D}{\Upsilon},\\
\tan\theta &= \frac{2g_D}{\Delta_D}.
\end{align}
\end{subequations}
It is then possible to write the block with respect to $\theta$:
\begin{align}
\Pi = \Upsilon\left[\begin{array}{cc}0&\frac{1}{2}\sin\theta\\\frac{1}{2}\sin\theta & \cos\theta\end{array}\right] = \Upsilon\left[\begin{array}{cc}0&\sin\frac{\theta}{2}\cos\frac{\theta}{2}\\\sin\frac{\theta}{2}\cos\frac{\theta}{2} & \cos^2\frac{\theta}{2} - \sin^2\frac{\theta}{2}\end{array}\right].
\end{align}
The eigenvalues can also be expressed in terms of the parametrized angle:
\begin{align}
\label{eigenv2}
\delta_\pm = \Upsilon\left\{\begin{array}{c}\cos^2\frac{\theta}{2}\\-\sin^2\frac{\theta}{2}\end{array}\right\},
\end{align}
which enables to write the transformation diagonalizing the block as:
\begin{align}
{\rm T}^\+\Pi{\rm T} = \Upsilon\left[\begin{array}{cc}-\sin^2\frac{\theta}{2} & 0\\0 & \cos^2\frac{\theta}{2}\end{array}\right].
\end{align}
Using the decomposition of the block $\Pi$ in terms of $\theta$, the unitary transformation ${\rm T}$ containing the eigenvectors $|\phi_\pm\ket$ associated with the eigenvalues $\delta_\pm$ reads:
\begin{align}
{\rm T} = \left[|\phi_-\ket,|\phi_+\ket\right] = \left[\begin{array}{cc} \cos\frac{\theta}{2}&\sin\frac{\theta}{2}\\-\sin\frac{\theta}{2}&\cos\frac{\theta}{2}\end{array}\right].
\end{align}
Once it is diagonalized, the $\Pi$ block is expressed in the basis of the polaritons $\{|\phi_-\ket,|\phi_+\ket\}$. The LP is associated with the subscript ($-$) while the UP is associated with the subscript (+).
  
  \subsection{Partial diagonalization of the 3$\times$3 Hamiltonian}

In this section we diagonalize partially the Hamiltonian \eqref{3x3} using the results of the previous section. To do so we create the following transformation:
\begin{align}
{\rm T}_3 = \left[\begin{array}{cc|c} \cos\frac{\theta}{2} & \sin\frac{\theta}{2} & 0\\-\sin\frac{\theta}{2} & \cos\frac{\theta}{2} & 0\\\hline 0 & 0 & 1\end{array}\right],
\end{align}
which transforms only the $\Pi$ block of the Hamiltonian. Changing the frame of reference of the Hamiltonian using this transformation, we get:
\begin{align}
\label{partH}
{\rm T}_3^\+{\rm H}{\rm T}_3 = \left[\begin{array}{cc|c}\delta_-&0&g_{BE}\cos\frac{\theta}{2}\\0&\delta_+&g_{BE}\sin\frac{\theta}{2}\\\hline g_B\cos\frac{\theta}{2}&g_B\sin\frac{\theta}{2}&\Delta_B\end{array}\right].
\end{align}
This Hamiltonian describes the interaction of both polaritons with a third state. Originally, only one of the polariton components is coupled to this state with coupling strength $g_B$, but the polaritons both couple to it with $g_B\cos\frac{\theta}{2}$ for ($-$) and $g_B\sin\frac{\theta}{2}$ for (+). Another consideration is how resonant the final system is. If the separation $\delta_+ - \delta_- = \Upsilon$ is larger than the linewidth of the third state, then only one polariton will couple efficiently with it. Finally, let's have a closer look at the sine and cosine factors. Using both \eqref{eigenv} and \eqref{eigenv2}, we find that these factors have the form:
\begin{subequations}
\label{sincos}
\begin{align}
\sin\frac{\theta}{2} &= \frac{1}{\sqrt{2}}\sqrt{1 - \frac{\Delta_D}{\sqrt{\Delta_D^2 + 4g_D^2}}}\\
\cos\frac{\theta}{2} &= \frac{1}{\sqrt{2}}\sqrt{1 + \frac{\Delta_D}{\sqrt{\Delta_D^2 + 4g_D^2}}}.
\end{align}
\end{subequations}
  \subsection{Optimal QE frequency and bright mode splitting}
In our system, the dark mode is located in the blue part of the spectrum. If the splitting between the dark mode and the QE is large enough, we expect the lower polariton to approach the resonance frequency of the bright mode and start interacting with it. If we look at the Hamiltonian in the partially diagonalized basis \eqref{partH} we see that the resonance happens for $\delta_- = \Delta_B$. We call the optimal QE-bright mode detuning $\Delta_B^{\rm opt}$ and using equations \eqref{eigenv} we find its value:
\begin{align}
\Delta_B^{\rm opt} = -\frac{g_D^2}{\Delta_D - \Delta_B} = -\frac{g_D^2}{\omega_D - \omega_B}.
\end{align}
The vacuum Rabi splitting of the bright mode is then calculated from equations \eqref{partH} and \eqref{sincos} and we get:
\begin{align}
\Omega_\text{bright} = 2g_B\cos\frac{\theta}{2} = \sqrt{2}g_B\sqrt{1 + \frac{\Delta_D}{\sqrt{\Delta_D^2 + 4g_D^2}}}.
\end{align}

\section{Master equation formalism and scattering spectrum}
  \subsection{Master equation in the weak pumping limit}
We use a master equation approach to calculate the scattering spectra in the main text. This approach not only corresponds to classical spectra in the weak pumping limit, but allows the modeling of quantum nonlinearities such as saturation of the bright mode that arise in the strong pumping limit. Here, we limit our study to the weak pumping limit and show the photon blockade by calculating the photon statistics of the scattered signal. The master equation corresponding to Hamiltonian \eqref{Hnheff} with a drive term is:
\begin{align}
\dot{\op{\varrho}} &= -i[\op{H},\op{\varrho}\,] + \gamma_B\Big(\op{a}_B\op{\varrho}\,\op{a}_B^\+ - \frac{1}{2}\op{\varrho}\,\op{a}_B^\+\op{a}_B - \frac{1}{2}\op{a}_B^\+\op{a}_B\op{\varrho}\Big) + \gamma_D\Big(\op{a}_D\op{\varrho}\,\op{a}_D^\+ - \frac{1}{2}\op{\varrho}\,\op{a}_D^\+\op{a}_D - \frac{1}{2}\op{a}_D^\+\op{a}_D\op{\varrho}\Big),\\
\op{H} &= \op{H}_S + \op{H}_\text{drive},\\
\op{H}_S &= \omega_E\hsig_+\hsig_- + \omega_B \op{a}_B^\+\op{a}_B + \omega_D\op{a}_D^\+\op{a}_D + g_B(\op{a}_B^\+\hsig_- + \op{a}_B\hsig_+) + g_D(\op{a}_D^\+\hsig_- + \op{a}_D\hsig_+),\\
\op{H}_\text{drive} &= -\Big(\frac{\mu_E}{\hbar}(\hsig_- + \hsig_+) + \frac{\mu_B}{\hbar}(\op{a}_B + \op{a}_B^\+)\Big)E_L\cos\omega_L t,
\end{align}
where $\op{\varrho}$ is the density operator for the emitter-bright mode-dark mode system, $\mu_{E,B}$ are the dipole moment of the emitter and the bright mode, respectively (we neglected the pumping term of the dark mode since it couples only locally to the QE), and the system is driven with a laser field amplitude $E_L$ and frequency $\omega_L$. Writing the Hamiltonian in the rotating frame of the driving field and applying the rotating wave approximation yields the Hamiltonian in the form:
\begin{multline}
\op{H} = \Delta_E\hsig_+\hsig_- + \Delta_B \op{a}_B^\+\op{a}_B + \Delta_D\op{a}_D^\+\op{a}_D + g_B(\op{a}_B^\+\hsig_- + \op{a}_B\hsig_+) + g_D(\op{a}_D^\+\hsig_- + \op{a}_D\hsig_+)
+ \frac{{\cal E}_E}{2}(\hsig_- + \hsig_+) + \frac{{\cal E}_B}{2}(\op{a}_B + \op{a}_B^\+),
\end{multline}
where $\Delta_j = \omega_j - \omega_L$ and ${\cal E}_j = -\mu_j E_L/\hbar$, $j = E,B,D$.
Since we study the weak pumping regime, the system is rarely in an excited state and thus the $\op{a}_{B,D}\op{\varrho}\,\op{a}_{B,D}^\+$ terms in the master equation can be neglected. This is equivalent to considering the effective Schr\"odinger equation:
\begin{align}
i\frac{\dd|\psi\ket}{\dd t} = \widetilde{H}|\psi\ket,
\end{align}
with $\widetilde{H} = \op{H} - i\frac{\gamma_E}{2}\hsig_+\hsig_- - i\frac{\gamma_B}{2}\op{a}_B^\+\op{a}_B - i\frac{\gamma_D}{2}\op{a}_D^\+\op{a}_D$ and whose steady-state solution yields the scattering spectrum and the photon statistics for zero delay. To solve this equation in the weak pumping limit, we proceed as in refs. \cite{SaezGarc17,CuarFern18} and solve for the steady-state:
\begin{subequations}
\begin{align}
|\psi_\text{s.s.}\ket &= \sum_{a = g,e}\sum_{b,c=0}^2c_{a,b,c}|a,b,c\ket\\
\widetilde{H}|\psi_\text{s.s.}\ket &= 0,
\end{align}
\end{subequations}
where we truncate the bright and dark excitation basis to 2, which is needed to evaluate the second order correlation function.

  \subsection{Scattering spectrum and second order correlation function}
The scattering spectrum is obtained by constructing the scattering operator:
\begin{align}
\op{a}_s = \mu_E\hsig_- + \mu_B\op{a}_B,
\end{align}
and computing the average over the steady-state of the associated number operator:
\begin{align}
S(\omega_L) = \bra\psi_\text{s.s.}|\op{a}_s^\+\op{a}_s|\psi_\text{s.s.}\ket.
\end{align}
When we consider the scattering map versus the nanosphere radius $R$, one should include the radius dependence of the scattered operator since larger nanospheres have larger dipole moments. To account for the radius dependence, we use the radiative decay rate formula from ref. \cite{Stockman11}:
%where we have introduced the radiative decay rate of the bright mode $\gamma_B^r$, which we calculate following ref. [Stockman]:
\begin{align}
\gamma_B^r = 4\varepsilon_b^{3/2}\left(\frac{\omega_{B}R}{c}\right)^3\left[\frac{\partial}{\partial\omega}\re\{\varepsilon_m(\omega)\}\right]^{-1}_{\omega=\omega_B},
\end{align}
$\varepsilon_b$ being the dielectric function of the surrounding medium, $\omega_{sp}$ being the surface plasmon resonance frequency of the nanoparticle (here considering Ag), $R$ being the radius of the nanoparticle and $\varepsilon_m(\omega)$ its Drude permittivity. The dipole moment $\mu_B$ is then given as a function of the radiative decay rate through the Fermi golden rule formula:
\begin{align}
\mu_B = \sqrt{\frac{3\hbar\pi\epsilon_b c^3}{\omega_B^3}\gamma_B^r}.
\end{align}
Finally, the second order correlation function for zero delay is given by the formula:
\begin{align}
g^{(2)}_{\omega_L}(0) = \frac{\bra\psi_\text{s.s.}|\op{a}_s^\+\op{a}_s^\+\op{a}_s\op{a}_s|\psi_\text{s.s.}\ket}{\bra\psi_\text{s.s.}|\op{a}_s^\+\op{a}_s|\psi_\text{s.s.}\ket^2}.
\end{align}

 \subsection{Scattering and photon statistics maps}
 \begin{figure*}
    \centering
    \includegraphics[width=\textwidth]{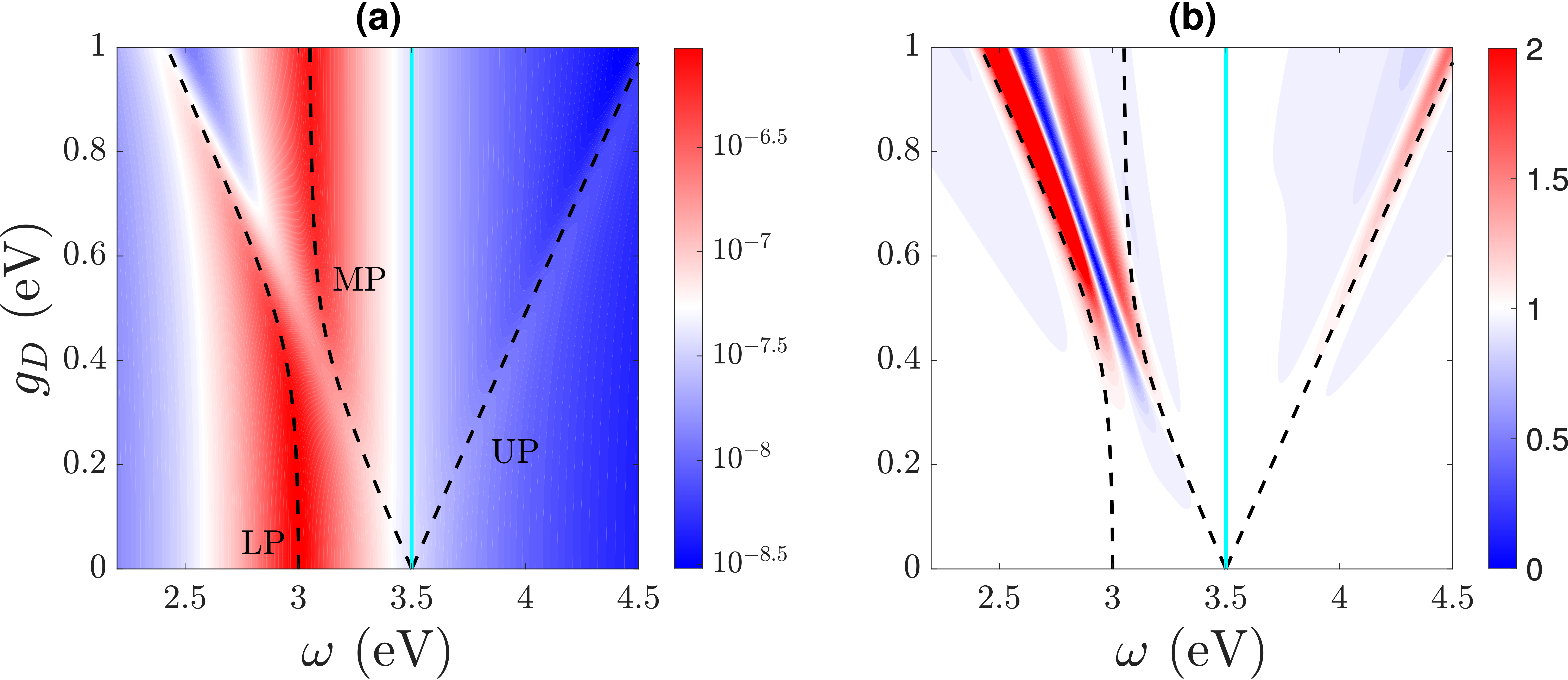}
    \caption{(a) Log-scale scattering intensity of the coupled QE-bright and dark mode system vs weak drive frequency $\omega$. Here $\omega_E = \omega_D = 3.5$ eV (light blue line), $\omega_B = 3$ eV and $g_D,g_B$ are linearly swept from 0 to 1 eV and 0 to 0.3 eV, respectively. (b) Zero-delay second order coherence function $g^{(2)}_\omega(0)$ with the same parametrization as in (a). Lower, middle and upper polariton (LP, MP, UP) lines are shown in dashed solid lines. Decay rates were fixed $\gamma_B = \gamma_D = 0.2$ eV and $\gamma_E = 0.1$ eV.}
    \label{dbdeltaB}
\end{figure*}
\begin{figure*}
    \centering
    \includegraphics[width=\textwidth]{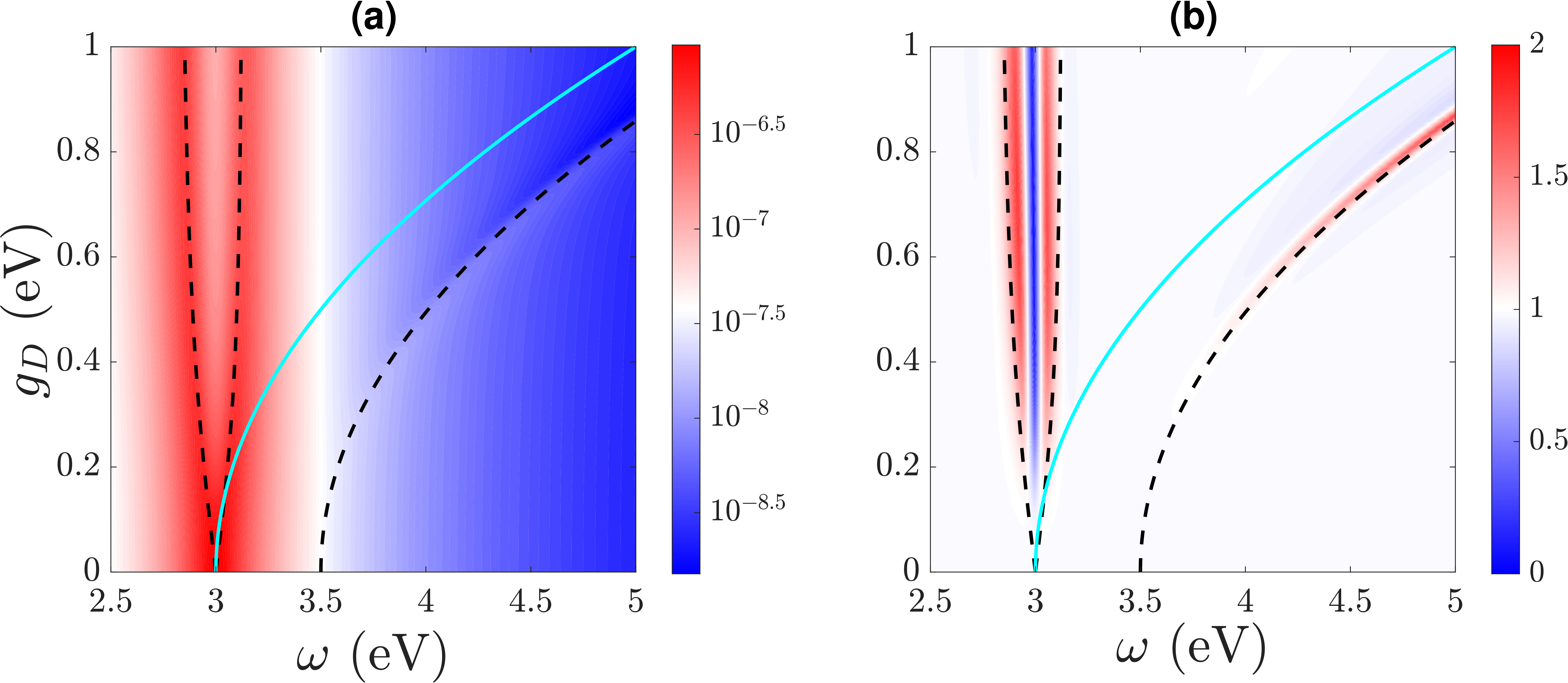}
    \caption{(a) Log-scale scattering intensity of the coupled QE-bright and dark mode system vs weak drive frequency $\omega$. All parameters are the same as in Fig. \ref{dbdeltaB} except for $\omega_E$, that is here taken to be the optimal frequency $\omega_E^\text{opt} = \omega_B + \delta_E^\text{opt}$. (b) Zero-delay second order coherence function $g^{(2)}_\omega(0)$ with the same parametrization as in (a). Lower, middle and upper polariton (LP, MP, UP) lines are shown in dashed solid lines.}
    \label{dbdeltaopt}
\end{figure*}
We finally discuss the scattering spectrum and the zero-delay second order coherence function of the system. In spite of Fig. 3 in the main text, corresponding to an Ag sphere coupled to a QE, here we model the Hamiltonian manually with similar parameters. Results are shown in Figs. \ref{dbdeltaB} and \ref{dbdeltaopt}. Fig. \ref{dbdeltaB}(a) shows the scattering when the QE frequency is tuned in resonance with the dark pseudomode $\omega_E = \omega_D$, and one can see an anticrossing between the LP and MP around $g_D = 0.5$ eV. The UP is here not very visible since it is strongly detuned with the bright mode. When the MP and LP are strongly coupled, an antibunching line appears in Fig. \ref{dbdeltaB}(b), and the latter is further red-shifted when $g_D$ and $g_B$ increase.
In Fig. \ref{dbdeltaopt} we plot the same data with the same parameters except for the QE frequency that is artificially swept in order to math the optimal frequency $\omega_E^\text{opt} = \omega_B + \delta_E^\text{opt}$, hence maintaining the optimal Rabi splitting between the LP and the MP. One can observe that for very high dark coupling strengths $g_D$ the UP and the emitter are very far detuned to the blue but the antibunching line in Fig. \ref{dbdeltaopt} (b) is kept constant.

\newpage
\twocolumngrid
\bibliography{main.bbl}

\begin{thebibliography}{52}%
\makeatletter
\providecommand \@ifxundefined [1]{%
 \@ifx{#1\undefined}
}%
\providecommand \@ifnum [1]{%
 \ifnum #1\expandafter \@firstoftwo
 \else \expandafter \@secondoftwo
 \fi
}%
\providecommand \@ifx [1]{%
 \ifx #1\expandafter \@firstoftwo
 \else \expandafter \@secondoftwo
 \fi
}%
\providecommand \natexlab [1]{#1}%
\providecommand \enquote  [1]{``#1''}%
\providecommand \bibnamefont  [1]{#1}%
\providecommand \bibfnamefont [1]{#1}%
\providecommand \citenamefont [1]{#1}%
\providecommand \href@noop [0]{\@secondoftwo}%
\providecommand \href [0]{\begingroup \@sanitize@url \@href}%
\providecommand \@href[1]{\@@startlink{#1}\@@href}%
\providecommand \@@href[1]{\endgroup#1\@@endlink}%
\providecommand \@sanitize@url [0]{\catcode `\\12\catcode `\$12\catcode
  `\&12\catcode `\#12\catcode `\^12\catcode `\_12\catcode `\%12\relax}%
\providecommand \@@startlink[1]{}%
\providecommand \@@endlink[0]{}%
\providecommand \url  [0]{\begingroup\@sanitize@url \@url }%
\providecommand \@url [1]{\endgroup\@href {#1}{\urlprefix }}%
\providecommand \urlprefix  [0]{URL }%
\providecommand \Eprint [0]{\href }%
\providecommand \doibase [0]{http://dx.doi.org/}%
\providecommand \selectlanguage [0]{\@gobble}%
\providecommand \bibinfo  [0]{\@secondoftwo}%
\providecommand \bibfield  [0]{\@secondoftwo}%
\providecommand \translation [1]{[#1]}%
\providecommand \BibitemOpen [0]{}%
\providecommand \bibitemStop [0]{}%
\providecommand \bibitemNoStop [0]{.\EOS\space}%
\providecommand \EOS [0]{\spacefactor3000\relax}%
\providecommand \BibitemShut  [1]{\csname bibitem#1\endcsname}%
\let\auto@bib@innerbib\@empty
%</preamble>
\bibitem [{\citenamefont {Scully}\ and\ \citenamefont
  {Zubairy}(1997)}]{Scully}%
  \BibitemOpen
  \bibfield  {author} {\bibinfo {author} {\bibfnamefont {M.~O.}\ \bibnamefont
  {Scully}}\ and\ \bibinfo {author} {\bibfnamefont {M.~S.}\ \bibnamefont
  {Zubairy}},\ }\href@noop {} {\emph {\bibinfo {title} {{Quantum optics}}}}\
  (\bibinfo  {publisher} {Cambridge University Press},\ \bibinfo {address}
  {Cambridge},\ \bibinfo {year} {1997})\ p.\ \bibinfo {pages} {656}\BibitemShut
  {NoStop}%
\bibitem [{\citenamefont {Tame}\ \emph {et~al.}(2013)\citenamefont {Tame},
  \citenamefont {McEnery}, \citenamefont {Ozdemir}, \citenamefont {Lee},
  \citenamefont {Maier},\ and\ \citenamefont {Kim}}]{Tame}%
  \BibitemOpen
  \bibfield  {author} {\bibinfo {author} {\bibfnamefont {M.~S.}\ \bibnamefont
  {Tame}}, \bibinfo {author} {\bibfnamefont {K.~R.}\ \bibnamefont {McEnery}},
  \bibinfo {author} {\bibfnamefont {S.~K.}\ \bibnamefont {Ozdemir}}, \bibinfo
  {author} {\bibfnamefont {J.}~\bibnamefont {Lee}}, \bibinfo {author}
  {\bibfnamefont {S.~A.}\ \bibnamefont {Maier}}, \ and\ \bibinfo {author}
  {\bibfnamefont {M.~S.}\ \bibnamefont {Kim}},\ }\href@noop {} {\bibfield
  {journal} {\bibinfo  {journal} {Nature Phys.}\ }\textbf {\bibinfo {volume}
  {9}},\ \bibinfo {pages} {329} (\bibinfo {year} {2013})}\BibitemShut {NoStop}%
\bibitem [{\citenamefont {Khitrova}\ \emph {et~al.}(2006)\citenamefont
  {Khitrova}, \citenamefont {Gibbs}, \citenamefont {Kira}, \citenamefont
  {Koch},\ and\ \citenamefont {Scherer}}]{Khitrova2006}%
  \BibitemOpen
  \bibfield  {author} {\bibinfo {author} {\bibfnamefont {G.}~\bibnamefont
  {Khitrova}}, \bibinfo {author} {\bibfnamefont {H.~M.}\ \bibnamefont {Gibbs}},
  \bibinfo {author} {\bibfnamefont {M.}~\bibnamefont {Kira}}, \bibinfo {author}
  {\bibfnamefont {S.~W.}\ \bibnamefont {Koch}}, \ and\ \bibinfo {author}
  {\bibfnamefont {A.}~\bibnamefont {Scherer}},\ }\href@noop {} {\bibfield
  {journal} {\bibinfo  {journal} {Nature Phys.}\ }\textbf {\bibinfo {volume}
  {2}},\ \bibinfo {pages} {81} (\bibinfo {year} {2006})}\BibitemShut {NoStop}%
\bibitem [{\citenamefont {Fink}\ \emph {et~al.}(2008)\citenamefont {Fink},
  \citenamefont {Goppl}, \citenamefont {Baur}, \citenamefont {Bianchetti},
  \citenamefont {Leek}, \citenamefont {Blais},\ and\ \citenamefont
  {Wallraff}}]{Fink2008}%
  \BibitemOpen
  \bibfield  {author} {\bibinfo {author} {\bibfnamefont {J.~M.}\ \bibnamefont
  {Fink}}, \bibinfo {author} {\bibfnamefont {M.}~\bibnamefont {Goppl}},
  \bibinfo {author} {\bibfnamefont {M.}~\bibnamefont {Baur}}, \bibinfo {author}
  {\bibfnamefont {R.}~\bibnamefont {Bianchetti}}, \bibinfo {author}
  {\bibfnamefont {P.~J.}\ \bibnamefont {Leek}}, \bibinfo {author}
  {\bibfnamefont {A.}~\bibnamefont {Blais}}, \ and\ \bibinfo {author}
  {\bibfnamefont {A.}~\bibnamefont {Wallraff}},\ }\href
  {http://dx.doi.org/10.1038/nature07112} {\bibfield  {journal} {\bibinfo
  {journal} {Nature}\ }\textbf {\bibinfo {volume} {454}},\ \bibinfo {pages}
  {315} (\bibinfo {year} {2008})}\BibitemShut {NoStop}%
\bibitem [{\citenamefont {T{\"{o}}rm{\"{a}}}\ and\ \citenamefont
  {Barnes}(2015)}]{Torma2015}%
  \BibitemOpen
  \bibfield  {author} {\bibinfo {author} {\bibfnamefont {P.}~\bibnamefont
  {T{\"{o}}rm{\"{a}}}}\ and\ \bibinfo {author} {\bibfnamefont {W.~L.}\
  \bibnamefont {Barnes}},\ }\href {http://www.ncbi.nlm.nih.gov/pubmed/25536670}
  {\bibfield  {journal} {\bibinfo  {journal} {Rep. Prog. Phys}\ }\textbf
  {\bibinfo {volume} {78}},\ \bibinfo {pages} {013901} (\bibinfo {year}
  {2015})}\BibitemShut {NoStop}%
\bibitem [{\citenamefont {Baranov}\ \emph {et~al.}(2018)\citenamefont
  {Baranov}, \citenamefont {Wersall}, \citenamefont {Cuadra}, \citenamefont
  {Antosiewicz},\ and\ \citenamefont {Shegai}}]{baranov2017novel}%
  \BibitemOpen
  \bibfield  {author} {\bibinfo {author} {\bibfnamefont {D.~G.}\ \bibnamefont
  {Baranov}}, \bibinfo {author} {\bibfnamefont {M.}~\bibnamefont {Wersall}},
  \bibinfo {author} {\bibfnamefont {J.}~\bibnamefont {Cuadra}}, \bibinfo
  {author} {\bibfnamefont {T.~J.}\ \bibnamefont {Antosiewicz}}, \ and\ \bibinfo
  {author} {\bibfnamefont {T.}~\bibnamefont {Shegai}},\ }\href@noop {}
  {\bibfield  {journal} {\bibinfo  {journal} {ACS Photonics}\ }\textbf
  {\bibinfo {volume} {5}},\ \bibinfo {pages} {24} (\bibinfo {year}
  {2018})}\BibitemShut {NoStop}%
\bibitem [{\citenamefont {Schachenmayer}\ \emph {et~al.}(2015)\citenamefont
  {Schachenmayer}, \citenamefont {Genes}, \citenamefont {Tignone},\ and\
  \citenamefont {Pupillo}}]{Transport}%
  \BibitemOpen
  \bibfield  {author} {\bibinfo {author} {\bibfnamefont {J.}~\bibnamefont
  {Schachenmayer}}, \bibinfo {author} {\bibfnamefont {C.}~\bibnamefont
  {Genes}}, \bibinfo {author} {\bibfnamefont {E.}~\bibnamefont {Tignone}}, \
  and\ \bibinfo {author} {\bibfnamefont {G.}~\bibnamefont {Pupillo}},\ }\href
  {http://link.aps.org/doi/10.1103/PhysRevLett.114.196403} {\bibfield
  {journal} {\bibinfo  {journal} {Phys. Rev. Lett.}\ }\textbf {\bibinfo
  {volume} {114}},\ \bibinfo {pages} {196403} (\bibinfo {year}
  {2015})}\BibitemShut {NoStop}%
\bibitem [{\citenamefont {Hutchison}\ \emph {et~al.}(2012)\citenamefont
  {Hutchison}, \citenamefont {Schwartz}, \citenamefont {Genet}, \citenamefont
  {Devaux},\ and\ \citenamefont {Ebbesen}}]{Angew2012}%
  \BibitemOpen
  \bibfield  {author} {\bibinfo {author} {\bibfnamefont {J.~A.}\ \bibnamefont
  {Hutchison}}, \bibinfo {author} {\bibfnamefont {T.}~\bibnamefont {Schwartz}},
  \bibinfo {author} {\bibfnamefont {C.}~\bibnamefont {Genet}}, \bibinfo
  {author} {\bibfnamefont {E.}~\bibnamefont {Devaux}}, \ and\ \bibinfo {author}
  {\bibfnamefont {T.~W.}\ \bibnamefont {Ebbesen}},\ }\href@noop {} {\bibfield
  {journal} {\bibinfo  {journal} {Angew. Chem., Int. Ed.}\ }\textbf {\bibinfo
  {volume} {51}},\ \bibinfo {pages} {1592} (\bibinfo {year}
  {2012})}\BibitemShut {NoStop}%
\bibitem [{\citenamefont {Thomas}\ \emph {et~al.}(2016)\citenamefont {Thomas},
  \citenamefont {George}, \citenamefont {Shalabney}, \citenamefont {Dryzhakov},
  \citenamefont {Varma}, \citenamefont {Moran}, \citenamefont {Chervy},
  \citenamefont {Xiaolan Zhong~and}, \citenamefont {Genet}, \citenamefont
  {Hutchison},\ and\ \citenamefont {Ebbesen}}]{Angew2016}%
  \BibitemOpen
  \bibfield  {author} {\bibinfo {author} {\bibfnamefont {A.}~\bibnamefont
  {Thomas}}, \bibinfo {author} {\bibfnamefont {J.}~\bibnamefont {George}},
  \bibinfo {author} {\bibfnamefont {A.}~\bibnamefont {Shalabney}}, \bibinfo
  {author} {\bibfnamefont {M.}~\bibnamefont {Dryzhakov}}, \bibinfo {author}
  {\bibfnamefont {S.~J.}\ \bibnamefont {Varma}}, \bibinfo {author}
  {\bibfnamefont {J.}~\bibnamefont {Moran}}, \bibinfo {author} {\bibfnamefont
  {T.}~\bibnamefont {Chervy}}, \bibinfo {author} {\bibfnamefont {E.~D.}\
  \bibnamefont {Xiaolan Zhong~and}}, \bibinfo {author} {\bibfnamefont
  {C.}~\bibnamefont {Genet}}, \bibinfo {author} {\bibfnamefont {J.~A.}\
  \bibnamefont {Hutchison}}, \ and\ \bibinfo {author} {\bibfnamefont {T.~W.}\
  \bibnamefont {Ebbesen}},\ }\href@noop {} {\bibfield  {journal} {\bibinfo
  {journal} {Angew. Chem. Int. Ed.}\ }\textbf {\bibinfo {volume} {55}},\
  \bibinfo {pages} {11462} (\bibinfo {year} {2016})}\BibitemShut {NoStop}%
\bibitem [{\citenamefont {Herrera}\ and\ \citenamefont
  {Spano}(2016)}]{Chemistry2016}%
  \BibitemOpen
  \bibfield  {author} {\bibinfo {author} {\bibfnamefont {F.}~\bibnamefont
  {Herrera}}\ and\ \bibinfo {author} {\bibfnamefont {F.~C.}\ \bibnamefont
  {Spano}},\ }\href {http://link.aps.org/doi/10.1103/PhysRevLett.116.238301}
  {\bibfield  {journal} {\bibinfo  {journal} {Phys. Rev. Lett.}\ }\textbf
  {\bibinfo {volume} {116}},\ \bibinfo {pages} {238301} (\bibinfo {year}
  {2016})}\BibitemShut {NoStop}%
\bibitem [{\citenamefont {Galego}\ \emph {et~al.}(2016)\citenamefont {Galego},
  \citenamefont {Garcia-Vidal},\ and\ \citenamefont {Feist}}]{Galego2016}%
  \BibitemOpen
  \bibfield  {author} {\bibinfo {author} {\bibfnamefont {J.}~\bibnamefont
  {Galego}}, \bibinfo {author} {\bibfnamefont {F.~J.}\ \bibnamefont
  {Garcia-Vidal}}, \ and\ \bibinfo {author} {\bibfnamefont {J.}~\bibnamefont
  {Feist}},\ }\href {http://arxiv.org/abs/1606.04684
  https://doi.org/10.1038/ncomms13841} {\bibfield  {journal} {\bibinfo
  {journal} {Nat. Commun.}\ }\textbf {\bibinfo {volume} {7}},\ \bibinfo {pages}
  {13841} (\bibinfo {year} {2016})}\BibitemShut {NoStop}%
\bibitem [{\citenamefont {Munkhbat}\ \emph {et~al.}(2018)\citenamefont
  {Munkhbat}, \citenamefont {Wers{\"a}ll}, \citenamefont {Baranov},
  \citenamefont {Antosiewicz},\ and\ \citenamefont
  {Shegai}}]{munkhbat2018suppression}%
  \BibitemOpen
  \bibfield  {author} {\bibinfo {author} {\bibfnamefont {B.}~\bibnamefont
  {Munkhbat}}, \bibinfo {author} {\bibfnamefont {M.}~\bibnamefont
  {Wers{\"a}ll}}, \bibinfo {author} {\bibfnamefont {D.~G.}\ \bibnamefont
  {Baranov}}, \bibinfo {author} {\bibfnamefont {T.~J.}\ \bibnamefont
  {Antosiewicz}}, \ and\ \bibinfo {author} {\bibfnamefont {T.}~\bibnamefont
  {Shegai}},\ }\href@noop {} {\bibfield  {journal} {\bibinfo  {journal} {Sci.
  Adv.}\ }\textbf {\bibinfo {volume} {4}},\ \bibinfo {pages} {eaas9552}
  (\bibinfo {year} {2018})}\BibitemShut {NoStop}%
\bibitem [{\citenamefont {Birnbaum}\ \emph {et~al.}(2005)\citenamefont
  {Birnbaum}, \citenamefont {Boca}, \citenamefont {Miller}, \citenamefont
  {Boozer}, \citenamefont {Northup},\ and\ \citenamefont
  {Kimble}}]{Birnbaum2005}%
  \BibitemOpen
  \bibfield  {author} {\bibinfo {author} {\bibfnamefont {K.~M.}\ \bibnamefont
  {Birnbaum}}, \bibinfo {author} {\bibfnamefont {A.}~\bibnamefont {Boca}},
  \bibinfo {author} {\bibfnamefont {R.}~\bibnamefont {Miller}}, \bibinfo
  {author} {\bibfnamefont {A.~D.}\ \bibnamefont {Boozer}}, \bibinfo {author}
  {\bibfnamefont {T.~E.}\ \bibnamefont {Northup}}, \ and\ \bibinfo {author}
  {\bibfnamefont {H.~J.}\ \bibnamefont {Kimble}},\ }\href
  {http://dx.doi.org/10.1038/nature03804} {\bibfield  {journal} {\bibinfo
  {journal} {Nature}\ }\textbf {\bibinfo {volume} {436}},\ \bibinfo {pages}
  {87} (\bibinfo {year} {2005})}\BibitemShut {NoStop}%
\bibitem [{\citenamefont {Englund}\ \emph {et~al.}(2007)\citenamefont
  {Englund}, \citenamefont {Faraon}, \citenamefont {Fushman}, \citenamefont
  {Stoltz}, \citenamefont {Petroff},\ and\ \citenamefont
  {Vuckovic}}]{Englund2007}%
  \BibitemOpen
  \bibfield  {author} {\bibinfo {author} {\bibfnamefont {D.}~\bibnamefont
  {Englund}}, \bibinfo {author} {\bibfnamefont {A.}~\bibnamefont {Faraon}},
  \bibinfo {author} {\bibfnamefont {I.}~\bibnamefont {Fushman}}, \bibinfo
  {author} {\bibfnamefont {N.}~\bibnamefont {Stoltz}}, \bibinfo {author}
  {\bibfnamefont {P.}~\bibnamefont {Petroff}}, \ and\ \bibinfo {author}
  {\bibfnamefont {J.}~\bibnamefont {Vuckovic}},\ }\href
  {http://dx.doi.org/10.1038/nature06234} {\bibfield  {journal} {\bibinfo
  {journal} {Nature}\ }\textbf {\bibinfo {volume} {450}},\ \bibinfo {pages}
  {857} (\bibinfo {year} {2007})}\BibitemShut {NoStop}%
\bibitem [{\citenamefont {Yoshie}\ \emph {et~al.}(2004)\citenamefont {Yoshie},
  \citenamefont {Scherer}, \citenamefont {Hendrickson}, \citenamefont
  {Khitrova}, \citenamefont {Gibbs}, \citenamefont {Rupper}, \citenamefont
  {Ell}, \citenamefont {Shchekin},\ and\ \citenamefont {Deppe}}]{Yoshie2004}%
  \BibitemOpen
  \bibfield  {author} {\bibinfo {author} {\bibfnamefont {T.}~\bibnamefont
  {Yoshie}}, \bibinfo {author} {\bibfnamefont {A.}~\bibnamefont {Scherer}},
  \bibinfo {author} {\bibfnamefont {J.}~\bibnamefont {Hendrickson}}, \bibinfo
  {author} {\bibfnamefont {G.}~\bibnamefont {Khitrova}}, \bibinfo {author}
  {\bibfnamefont {H.~M.}\ \bibnamefont {Gibbs}}, \bibinfo {author}
  {\bibfnamefont {G.}~\bibnamefont {Rupper}}, \bibinfo {author} {\bibfnamefont
  {C.}~\bibnamefont {Ell}}, \bibinfo {author} {\bibfnamefont {O.~B.}\
  \bibnamefont {Shchekin}}, \ and\ \bibinfo {author} {\bibfnamefont {D.~G.}\
  \bibnamefont {Deppe}},\ }\href {http://www.ncbi.nlm.nih.gov/pubmed/15538363}
  {\bibfield  {journal} {\bibinfo  {journal} {Nature}\ }\textbf {\bibinfo
  {volume} {432}},\ \bibinfo {pages} {9} (\bibinfo {year} {2004})}\BibitemShut
  {NoStop}%
\bibitem [{\citenamefont {Reithmaier}\ \emph {et~al.}(2004)\citenamefont
  {Reithmaier}, \citenamefont {Sek}, \citenamefont {L{\"{o}}ffler},
  \citenamefont {Hofmann}, \citenamefont {Kuhn}, \citenamefont {Reitzenstein},
  \citenamefont {Keldysh}, \citenamefont {Kulakovskii}, \citenamefont
  {Reinecke},\ and\ \citenamefont {Forchel}}]{Reithmaier2004}%
  \BibitemOpen
  \bibfield  {author} {\bibinfo {author} {\bibfnamefont {J.~P.}\ \bibnamefont
  {Reithmaier}}, \bibinfo {author} {\bibfnamefont {G.}~\bibnamefont {Sek}},
  \bibinfo {author} {\bibfnamefont {A.}~\bibnamefont {L{\"{o}}ffler}}, \bibinfo
  {author} {\bibfnamefont {C.}~\bibnamefont {Hofmann}}, \bibinfo {author}
  {\bibfnamefont {S.}~\bibnamefont {Kuhn}}, \bibinfo {author} {\bibfnamefont
  {S.}~\bibnamefont {Reitzenstein}}, \bibinfo {author} {\bibfnamefont {L.~V.}\
  \bibnamefont {Keldysh}}, \bibinfo {author} {\bibfnamefont {V.~D.}\
  \bibnamefont {Kulakovskii}}, \bibinfo {author} {\bibfnamefont {T.~L.}\
  \bibnamefont {Reinecke}}, \ and\ \bibinfo {author} {\bibfnamefont
  {A.}~\bibnamefont {Forchel}},\ }\href {http://dx.doi.org/10.1038/nature02969}
  {\bibfield  {journal} {\bibinfo  {journal} {Nature}\ }\textbf {\bibinfo
  {volume} {432}},\ \bibinfo {pages} {197} (\bibinfo {year}
  {2004})}\BibitemShut {NoStop}%
\bibitem [{\citenamefont {Chikkaraddy}\ \emph {et~al.}(2016)\citenamefont
  {Chikkaraddy}, \citenamefont {de~Nijs}, \citenamefont {Benz}, \citenamefont
  {Barrow}, \citenamefont {Scherman}, \citenamefont {Rosta}, \citenamefont
  {Demetriadou}, \citenamefont {Fox}, \citenamefont {Hess},\ and\ \citenamefont
  {Baumberg}}]{Chikkaraddy2016}%
  \BibitemOpen
  \bibfield  {author} {\bibinfo {author} {\bibfnamefont {R.}~\bibnamefont
  {Chikkaraddy}}, \bibinfo {author} {\bibfnamefont {B.}~\bibnamefont
  {de~Nijs}}, \bibinfo {author} {\bibfnamefont {F.}~\bibnamefont {Benz}},
  \bibinfo {author} {\bibfnamefont {S.~J.}\ \bibnamefont {Barrow}}, \bibinfo
  {author} {\bibfnamefont {O.~A.}\ \bibnamefont {Scherman}}, \bibinfo {author}
  {\bibfnamefont {E.}~\bibnamefont {Rosta}}, \bibinfo {author} {\bibfnamefont
  {A.}~\bibnamefont {Demetriadou}}, \bibinfo {author} {\bibfnamefont
  {P.}~\bibnamefont {Fox}}, \bibinfo {author} {\bibfnamefont {O.}~\bibnamefont
  {Hess}}, \ and\ \bibinfo {author} {\bibfnamefont {J.~J.}\ \bibnamefont
  {Baumberg}},\ }\href {http://www.nature.com/doifinder/10.1038/nature17974}
  {\bibfield  {journal} {\bibinfo  {journal} {Nature}\ }\textbf {\bibinfo
  {volume} {535}},\ \bibinfo {pages} {127} (\bibinfo {year}
  {2016})}\BibitemShut {NoStop}%
\bibitem [{\citenamefont {Santhosh}\ \emph {et~al.}(2016)\citenamefont
  {Santhosh}, \citenamefont {Bitton}, \citenamefont {Chuntonov},\ and\
  \citenamefont {Haran}}]{Santhosh2016}%
  \BibitemOpen
  \bibfield  {author} {\bibinfo {author} {\bibfnamefont {K.}~\bibnamefont
  {Santhosh}}, \bibinfo {author} {\bibfnamefont {O.}~\bibnamefont {Bitton}},
  \bibinfo {author} {\bibfnamefont {L.}~\bibnamefont {Chuntonov}}, \ and\
  \bibinfo {author} {\bibfnamefont {G.}~\bibnamefont {Haran}},\ }\href
  {http://arxiv.org/abs/1511.00263} {\bibfield  {journal} {\bibinfo  {journal}
  {Nat. Commun.}\ }\textbf {\bibinfo {volume} {7}},\ \bibinfo {pages} {11823}
  (\bibinfo {year} {2016})}\BibitemShut {NoStop}%
\bibitem [{\citenamefont {Gro{\ss}}\ \emph {et~al.}(2018)\citenamefont
  {Gro{\ss}}, \citenamefont {Hamm}, \citenamefont {Tufarelli}, \citenamefont
  {Hess},\ and\ \citenamefont {Hecht}}]{gross2018near}%
  \BibitemOpen
  \bibfield  {author} {\bibinfo {author} {\bibfnamefont {H.}~\bibnamefont
  {Gro{\ss}}}, \bibinfo {author} {\bibfnamefont {J.~M.}\ \bibnamefont {Hamm}},
  \bibinfo {author} {\bibfnamefont {T.}~\bibnamefont {Tufarelli}}, \bibinfo
  {author} {\bibfnamefont {O.}~\bibnamefont {Hess}}, \ and\ \bibinfo {author}
  {\bibfnamefont {B.}~\bibnamefont {Hecht}},\ }\href@noop {} {\bibfield
  {journal} {\bibinfo  {journal} {Sci. Adv.}\ }\textbf {\bibinfo {volume}
  {4}},\ \bibinfo {pages} {eaar4906} (\bibinfo {year} {2018})}\BibitemShut
  {NoStop}%
\bibitem [{\citenamefont {Leng}\ \emph {et~al.}(2018)\citenamefont {Leng},
  \citenamefont {Szychowski}, \citenamefont {Daniel},\ and\ \citenamefont
  {Pelton}}]{PeltonNat18}%
  \BibitemOpen
  \bibfield  {author} {\bibinfo {author} {\bibfnamefont {H.}~\bibnamefont
  {Leng}}, \bibinfo {author} {\bibfnamefont {B.}~\bibnamefont {Szychowski}},
  \bibinfo {author} {\bibfnamefont {M.-C.}\ \bibnamefont {Daniel}}, \ and\
  \bibinfo {author} {\bibfnamefont {M.}~\bibnamefont {Pelton}},\ }\href
  {\doibase 10.1038/s41467-018-06450-4} {\bibfield  {journal} {\bibinfo
  {journal} {Nature Communications}\ }\textbf {\bibinfo {volume} {9}},\
  \bibinfo {pages} {4012} (\bibinfo {year} {2018})}\BibitemShut {NoStop}%
\bibitem [{\citenamefont {Rousseaux}\ \emph {et~al.}(2018)\citenamefont
  {Rousseaux}, \citenamefont {Baranov}, \citenamefont {K\"all}, \citenamefont
  {Shegai},\ and\ \citenamefont {Johansson}}]{rousseaux2018comparative}%
  \BibitemOpen
  \bibfield  {author} {\bibinfo {author} {\bibfnamefont {B.}~\bibnamefont
  {Rousseaux}}, \bibinfo {author} {\bibfnamefont {D.~G.}\ \bibnamefont
  {Baranov}}, \bibinfo {author} {\bibfnamefont {M.}~\bibnamefont {K\"all}},
  \bibinfo {author} {\bibfnamefont {T.}~\bibnamefont {Shegai}}, \ and\ \bibinfo
  {author} {\bibfnamefont {G.}~\bibnamefont {Johansson}},\ }\href {\doibase
  10.1103/PhysRevB.98.045435} {\bibfield  {journal} {\bibinfo  {journal} {Phys.
  Rev. B}\ }\textbf {\bibinfo {volume} {98}},\ \bibinfo {pages} {045435}
  (\bibinfo {year} {2018})}\BibitemShut {NoStop}%
\bibitem [{\citenamefont {Leatherdale}\ \emph
  {et~al.}(2002{\natexlab{a}})\citenamefont {Leatherdale}, \citenamefont {Woo},
  \citenamefont {Mikulec},\ and\ \citenamefont
  {Bawendi}}]{leatherdale2002absorption}%
  \BibitemOpen
  \bibfield  {author} {\bibinfo {author} {\bibfnamefont {C.~A.}\ \bibnamefont
  {Leatherdale}}, \bibinfo {author} {\bibfnamefont {W.-K.}\ \bibnamefont
  {Woo}}, \bibinfo {author} {\bibfnamefont {F.~V.}\ \bibnamefont {Mikulec}}, \
  and\ \bibinfo {author} {\bibfnamefont {M.~G.}\ \bibnamefont {Bawendi}},\
  }\href@noop {} {\bibfield  {journal} {\bibinfo  {journal} {J. Phys. Chem. B}\
  }\textbf {\bibinfo {volume} {106}},\ \bibinfo {pages} {7619} (\bibinfo {year}
  {2002}{\natexlab{a}})}\BibitemShut {NoStop}%
\bibitem [{\citenamefont {Yu}\ \emph {et~al.}(2003{\natexlab{a}})\citenamefont
  {Yu}, \citenamefont {Qu}, \citenamefont {Guo},\ and\ \citenamefont
  {Peng}}]{QDreview2003}%
  \BibitemOpen
  \bibfield  {author} {\bibinfo {author} {\bibfnamefont {W.~W.}\ \bibnamefont
  {Yu}}, \bibinfo {author} {\bibfnamefont {L.}~\bibnamefont {Qu}}, \bibinfo
  {author} {\bibfnamefont {W.}~\bibnamefont {Guo}}, \ and\ \bibinfo {author}
  {\bibfnamefont {X.}~\bibnamefont {Peng}},\ }\href {\doibase
  10.1021/cm034081k} {\bibfield  {journal} {\bibinfo  {journal} {Chemistry of
  Materials}\ }\textbf {\bibinfo {volume} {15}},\ \bibinfo {pages} {2854}
  (\bibinfo {year} {2003}{\natexlab{a}})},\ \Eprint
  {http://arxiv.org/abs/https://doi.org/10.1021/cm034081k}
  {https://doi.org/10.1021/cm034081k} \BibitemShut {NoStop}%
\bibitem [{\citenamefont {Wang}\ \emph {et~al.}(2017)\citenamefont {Wang},
  \citenamefont {Chernikov}, \citenamefont {Glazov}, \citenamefont {Heinz},
  \citenamefont {Marie}, \citenamefont {Amand},\ and\ \citenamefont
  {Urbaszek}}]{wang2017excitons}%
  \BibitemOpen
  \bibfield  {author} {\bibinfo {author} {\bibfnamefont {G.}~\bibnamefont
  {Wang}}, \bibinfo {author} {\bibfnamefont {A.}~\bibnamefont {Chernikov}},
  \bibinfo {author} {\bibfnamefont {M.~M.}\ \bibnamefont {Glazov}}, \bibinfo
  {author} {\bibfnamefont {T.~F.}\ \bibnamefont {Heinz}}, \bibinfo {author}
  {\bibfnamefont {X.}~\bibnamefont {Marie}}, \bibinfo {author} {\bibfnamefont
  {T.}~\bibnamefont {Amand}}, \ and\ \bibinfo {author} {\bibfnamefont
  {B.}~\bibnamefont {Urbaszek}},\ }\href@noop {} {\bibfield  {journal}
  {\bibinfo  {journal} {arXiv preprint arXiv:1707.05863}\ } (\bibinfo {year}
  {2017})}\BibitemShut {NoStop}%
\bibitem [{\citenamefont {Knight}\ \emph {et~al.}(2013)\citenamefont {Knight},
  \citenamefont {King}, \citenamefont {Liu}, \citenamefont {Everitt},
  \citenamefont {Nordlander},\ and\ \citenamefont
  {Halas}}]{knight2013aluminum}%
  \BibitemOpen
  \bibfield  {author} {\bibinfo {author} {\bibfnamefont {M.~W.}\ \bibnamefont
  {Knight}}, \bibinfo {author} {\bibfnamefont {N.~S.}\ \bibnamefont {King}},
  \bibinfo {author} {\bibfnamefont {L.}~\bibnamefont {Liu}}, \bibinfo {author}
  {\bibfnamefont {H.~O.}\ \bibnamefont {Everitt}}, \bibinfo {author}
  {\bibfnamefont {P.}~\bibnamefont {Nordlander}}, \ and\ \bibinfo {author}
  {\bibfnamefont {N.~J.}\ \bibnamefont {Halas}},\ }\href@noop {} {\bibfield
  {journal} {\bibinfo  {journal} {ACS Nano}\ }\textbf {\bibinfo {volume} {8}},\
  \bibinfo {pages} {834} (\bibinfo {year} {2013})}\BibitemShut {NoStop}%
\bibitem [{\citenamefont {Rossi}\ \emph {et~al.}(2019)\citenamefont {Rossi},
  \citenamefont {Shegai}, \citenamefont {Erhart},\ and\ \citenamefont
  {Antosiewicz}}]{rossi2019}%
  \BibitemOpen
  \bibfield  {author} {\bibinfo {author} {\bibfnamefont {T.~P.}\ \bibnamefont
  {Rossi}}, \bibinfo {author} {\bibfnamefont {T.}~\bibnamefont {Shegai}},
  \bibinfo {author} {\bibfnamefont {P.}~\bibnamefont {Erhart}}, \ and\ \bibinfo
  {author} {\bibfnamefont {T.~J.}\ \bibnamefont {Antosiewicz}},\ }\href@noop {}
  {\bibfield  {journal} {\bibinfo  {journal} {Nature Commun.}\ }\textbf
  {\bibinfo {volume} {10}},\ \bibinfo {pages} {3336} (\bibinfo {year}
  {2019})}\BibitemShut {NoStop}%
\bibitem [{\citenamefont {Liu}\ \emph {et~al.}(2009)\citenamefont {Liu},
  \citenamefont {Lee}, \citenamefont {Gray}, \citenamefont {Guyot-Sionnest},
  \citenamefont {Pelton} \emph {et~al.}}]{liu2009excitation}%
  \BibitemOpen
  \bibfield  {author} {\bibinfo {author} {\bibfnamefont {M.}~\bibnamefont
  {Liu}}, \bibinfo {author} {\bibfnamefont {T.-W.}\ \bibnamefont {Lee}},
  \bibinfo {author} {\bibfnamefont {S.~K.}\ \bibnamefont {Gray}}, \bibinfo
  {author} {\bibfnamefont {P.}~\bibnamefont {Guyot-Sionnest}}, \bibinfo
  {author} {\bibfnamefont {M.}~\bibnamefont {Pelton}},  \emph {et~al.},\
  }\href@noop {} {\bibfield  {journal} {\bibinfo  {journal} {Physical review
  letters}\ }\textbf {\bibinfo {volume} {102}},\ \bibinfo {pages} {107401}
  (\bibinfo {year} {2009})}\BibitemShut {NoStop}%
\bibitem [{\citenamefont {Delga}\ \emph {et~al.}(2014)\citenamefont {Delga},
  \citenamefont {Feist}, \citenamefont {Bravo-Abad},\ and\ \citenamefont
  {Garcia-Vidal}}]{Delga_2014}%
  \BibitemOpen
  \bibfield  {author} {\bibinfo {author} {\bibfnamefont {A.}~\bibnamefont
  {Delga}}, \bibinfo {author} {\bibfnamefont {J.}~\bibnamefont {Feist}},
  \bibinfo {author} {\bibfnamefont {J.}~\bibnamefont {Bravo-Abad}}, \ and\
  \bibinfo {author} {\bibfnamefont {F.~J.}\ \bibnamefont {Garcia-Vidal}},\
  }\href@noop {} {\bibfield  {journal} {\bibinfo  {journal} {Phys. Rev. Lett.}\
  }\textbf {\bibinfo {volume} {112}},\ \bibinfo {pages} {253601} (\bibinfo
  {year} {2014})}\BibitemShut {NoStop}%
\bibitem [{\citenamefont {Rousseaux}\ \emph {et~al.}(2016)\citenamefont
  {Rousseaux}, \citenamefont {Dzsotjan}, \citenamefont {Colas~des Francs},
  \citenamefont {Jauslin}, \citenamefont {Couteau},\ and\ \citenamefont
  {Gu\'erin}}]{RousGuer16}%
  \BibitemOpen
  \bibfield  {author} {\bibinfo {author} {\bibfnamefont {B.}~\bibnamefont
  {Rousseaux}}, \bibinfo {author} {\bibfnamefont {D.}~\bibnamefont {Dzsotjan}},
  \bibinfo {author} {\bibfnamefont {G.}~\bibnamefont {Colas~des Francs}},
  \bibinfo {author} {\bibfnamefont {H.~R.}\ \bibnamefont {Jauslin}}, \bibinfo
  {author} {\bibfnamefont {C.}~\bibnamefont {Couteau}}, \ and\ \bibinfo
  {author} {\bibfnamefont {S.}~\bibnamefont {Gu\'erin}},\ }\href {\doibase
  10.1103/PhysRevB.93.045422} {\bibfield  {journal} {\bibinfo  {journal} {Phys.
  Rev. B}\ }\textbf {\bibinfo {volume} {93}},\ \bibinfo {pages} {045422}
  (\bibinfo {year} {2016})}\BibitemShut {NoStop}%
\bibitem [{\citenamefont {Varguet}\ \emph {et~al.}(2016)\citenamefont
  {Varguet}, \citenamefont {Rousseaux}, \citenamefont {Dzsotjan}, \citenamefont
  {Jauslin}, \citenamefont {Guérin},\ and\ \citenamefont {des
  Francs}}]{VargCola16}%
  \BibitemOpen
  \bibfield  {author} {\bibinfo {author} {\bibfnamefont {H.}~\bibnamefont
  {Varguet}}, \bibinfo {author} {\bibfnamefont {B.}~\bibnamefont {Rousseaux}},
  \bibinfo {author} {\bibfnamefont {D.}~\bibnamefont {Dzsotjan}}, \bibinfo
  {author} {\bibfnamefont {H.-R.}\ \bibnamefont {Jauslin}}, \bibinfo {author}
  {\bibfnamefont {S.}~\bibnamefont {Guérin}}, \ and\ \bibinfo {author}
  {\bibfnamefont {G.~C.}\ \bibnamefont {des Francs}},\ }\href@noop {}
  {\bibfield  {journal} {\bibinfo  {journal} {Opt. Lett.}\ }\textbf {\bibinfo
  {volume} {41}},\ \bibinfo {pages} {4480} (\bibinfo {year}
  {2016})}\BibitemShut {NoStop}%
\bibitem [{\citenamefont {Li}\ \emph {et~al.}(2018)\citenamefont {Li},
  \citenamefont {Garc{\'\i}a-Vidal},\ and\ \citenamefont
  {Fern{\'a}ndez-Dom{\'\i}nguez}}]{LiFern18}%
  \BibitemOpen
  \bibfield  {author} {\bibinfo {author} {\bibfnamefont {R.-Q.}\ \bibnamefont
  {Li}}, \bibinfo {author} {\bibfnamefont {F.~J.}\ \bibnamefont
  {Garc{\'\i}a-Vidal}}, \ and\ \bibinfo {author} {\bibfnamefont {A.~I.}\
  \bibnamefont {Fern{\'a}ndez-Dom{\'\i}nguez}},\ }\bibfield  {booktitle} {\emph
  {\bibinfo {booktitle} {ACS Photonics}},\ }\href {\doibase
  10.1021/acsphotonics.7b00616} {\bibfield  {journal} {\bibinfo  {journal} {ACS
  Photonics}\ }\textbf {\bibinfo {volume} {5}},\ \bibinfo {pages} {177}
  (\bibinfo {year} {2018})}\BibitemShut {NoStop}%
\bibitem [{\citenamefont {Castellini}\ \emph {et~al.}(2018)\citenamefont
  {Castellini}, \citenamefont {Jauslin}, \citenamefont {Rousseaux},
  \citenamefont {Dzsotjan}, \citenamefont {Colas~des Francs}, \citenamefont
  {Messina},\ and\ \citenamefont {Gu{\'e}rin}}]{CastGuer18}%
  \BibitemOpen
  \bibfield  {author} {\bibinfo {author} {\bibfnamefont {A.}~\bibnamefont
  {Castellini}}, \bibinfo {author} {\bibfnamefont {H.~R.}\ \bibnamefont
  {Jauslin}}, \bibinfo {author} {\bibfnamefont {B.}~\bibnamefont {Rousseaux}},
  \bibinfo {author} {\bibfnamefont {D.}~\bibnamefont {Dzsotjan}}, \bibinfo
  {author} {\bibfnamefont {G.}~\bibnamefont {Colas~des Francs}}, \bibinfo
  {author} {\bibfnamefont {A.}~\bibnamefont {Messina}}, \ and\ \bibinfo
  {author} {\bibfnamefont {S.}~\bibnamefont {Gu{\'e}rin}},\ }\href {\doibase
  10.1140/epjd/e2018-90322-5} {\bibfield  {journal} {\bibinfo  {journal} {The
  European Physical Journal D}\ }\textbf {\bibinfo {volume} {72}},\ \bibinfo
  {pages} {223} (\bibinfo {year} {2018})}\BibitemShut {NoStop}%
\bibitem [{\citenamefont {Varguet}\ \emph {et~al.}(2019)\citenamefont
  {Varguet}, \citenamefont {Rousseaux}, \citenamefont {Dzsotjan}, \citenamefont
  {Jauslin}, \citenamefont {Gu{\'{e}}rin},\ and\ \citenamefont {des
  Francs}}]{VargCola19}%
  \BibitemOpen
  \bibfield  {author} {\bibinfo {author} {\bibfnamefont {H.}~\bibnamefont
  {Varguet}}, \bibinfo {author} {\bibfnamefont {B.}~\bibnamefont {Rousseaux}},
  \bibinfo {author} {\bibfnamefont {D.}~\bibnamefont {Dzsotjan}}, \bibinfo
  {author} {\bibfnamefont {H.~R.}\ \bibnamefont {Jauslin}}, \bibinfo {author}
  {\bibfnamefont {S.}~\bibnamefont {Gu{\'{e}}rin}}, \ and\ \bibinfo {author}
  {\bibfnamefont {G.~C.}\ \bibnamefont {des Francs}},\ }\href {\doibase
  10.1088/1361-6455/ab008e} {\bibfield  {journal} {\bibinfo  {journal} {Journal
  of Physics B: Atomic, Molecular and Optical Physics}\ }\textbf {\bibinfo
  {volume} {52}},\ \bibinfo {pages} {055404} (\bibinfo {year}
  {2019})}\BibitemShut {NoStop}%
\bibitem [{\citenamefont {Cuartero-Gonz\'alez}\ and\ \citenamefont
  {Fern\'andez-Dom\'inguez}()}]{CuarFernPre}%
  \BibitemOpen
  \bibfield  {author} {\bibinfo {author} {\bibfnamefont {A.}~\bibnamefont
  {Cuartero-Gonz\'alez}}\ and\ \bibinfo {author} {\bibfnamefont {A.~I.}\
  \bibnamefont {Fern\'andez-Dom\'inguez}},\ }\href
  {https://arxiv.org/abs/1905.09893} {\bibinfo  {journal} {arXiv:1905.09893}\
  }\BibitemShut {NoStop}%
\bibitem [{\citenamefont {Koh}\ \emph {et~al.}(2009)\citenamefont {Koh},
  \citenamefont {Bao}, \citenamefont {Khan}, \citenamefont {Smith},
  \citenamefont {Kothleitner}, \citenamefont {Nordlander}, \citenamefont
  {Maier},\ and\ \citenamefont {McComb}}]{koh2009electron}%
  \BibitemOpen
\bibfield  {journal} {  }\bibfield  {author} {\bibinfo {author} {\bibfnamefont
  {A.~L.}\ \bibnamefont {Koh}}, \bibinfo {author} {\bibfnamefont
  {K.}~\bibnamefont {Bao}}, \bibinfo {author} {\bibfnamefont {I.}~\bibnamefont
  {Khan}}, \bibinfo {author} {\bibfnamefont {W.~E.}\ \bibnamefont {Smith}},
  \bibinfo {author} {\bibfnamefont {G.}~\bibnamefont {Kothleitner}}, \bibinfo
  {author} {\bibfnamefont {P.}~\bibnamefont {Nordlander}}, \bibinfo {author}
  {\bibfnamefont {S.~A.}\ \bibnamefont {Maier}}, \ and\ \bibinfo {author}
  {\bibfnamefont {D.~W.}\ \bibnamefont {McComb}},\ }\href@noop {} {\bibfield
  {journal} {\bibinfo  {journal} {ACS Nano}\ }\textbf {\bibinfo {volume} {3}},\
  \bibinfo {pages} {3015} (\bibinfo {year} {2009})}\BibitemShut {NoStop}%
\bibitem [{\citenamefont {Barrow}\ \emph {et~al.}(2014)\citenamefont {Barrow},
  \citenamefont {Rossouw}, \citenamefont {Funston}, \citenamefont {Botton},\
  and\ \citenamefont {Mulvaney}}]{barrow2014mapping}%
  \BibitemOpen
  \bibfield  {author} {\bibinfo {author} {\bibfnamefont {S.~J.}\ \bibnamefont
  {Barrow}}, \bibinfo {author} {\bibfnamefont {D.}~\bibnamefont {Rossouw}},
  \bibinfo {author} {\bibfnamefont {A.~M.}\ \bibnamefont {Funston}}, \bibinfo
  {author} {\bibfnamefont {G.~A.}\ \bibnamefont {Botton}}, \ and\ \bibinfo
  {author} {\bibfnamefont {P.}~\bibnamefont {Mulvaney}},\ }\href@noop {}
  {\bibfield  {journal} {\bibinfo  {journal} {Nano Lett.}\ }\textbf {\bibinfo
  {volume} {14}},\ \bibinfo {pages} {3799} (\bibinfo {year}
  {2014})}\BibitemShut {NoStop}%
\bibitem [{\citenamefont {Bitton}\ \emph {et~al.}()\citenamefont {Bitton},
  \citenamefont {Gupta}, \citenamefont {Houben}, \citenamefont {Kvapil},
  \citenamefont {Krapek}, \citenamefont {Sikola},\ and\ \citenamefont
  {Haran}}]{BittHaraPre}%
  \BibitemOpen
  \bibfield  {author} {\bibinfo {author} {\bibfnamefont {O.}~\bibnamefont
  {Bitton}}, \bibinfo {author} {\bibfnamefont {S.~N.}\ \bibnamefont {Gupta}},
  \bibinfo {author} {\bibfnamefont {L.}~\bibnamefont {Houben}}, \bibinfo
  {author} {\bibfnamefont {M.}~\bibnamefont {Kvapil}}, \bibinfo {author}
  {\bibfnamefont {V.}~\bibnamefont {Krapek}}, \bibinfo {author} {\bibfnamefont
  {T.}~\bibnamefont {Sikola}}, \ and\ \bibinfo {author} {\bibfnamefont
  {G.}~\bibnamefont {Haran}},\ }\href {https://arxiv.org/abs/1907.10299}
  {\bibinfo  {journal} {arXiv:1907.10299}\ }\BibitemShut {NoStop}%
\bibitem [{\citenamefont {Novotny}\ and\ \citenamefont
  {Hecht}(2012)}]{LukasNovotny2012}%
  \BibitemOpen
\bibfield  {journal} {  }\bibfield  {author} {\bibinfo {author} {\bibfnamefont
  {L.}~\bibnamefont {Novotny}}\ and\ \bibinfo {author} {\bibfnamefont
  {B.}~\bibnamefont {Hecht}},\ }\href@noop {} {\emph {\bibinfo {title}
  {{Principles of Nano-Optics}}}}\ (\bibinfo  {publisher} {Cambridge University
  Press},\ \bibinfo {year} {2012})\BibitemShut {NoStop}%
\bibitem [{\citenamefont {Anger}\ \emph {et~al.}(2006)\citenamefont {Anger},
  \citenamefont {Bharadwaj},\ and\ \citenamefont {Novotny}}]{Anger2006}%
  \BibitemOpen
  \bibfield  {author} {\bibinfo {author} {\bibfnamefont {P.}~\bibnamefont
  {Anger}}, \bibinfo {author} {\bibfnamefont {P.}~\bibnamefont {Bharadwaj}}, \
  and\ \bibinfo {author} {\bibfnamefont {L.}~\bibnamefont {Novotny}},\ }\href
  {http://link.aps.org/doi/10.1103/PhysRevLett.96.113002} {\bibfield  {journal}
  {\bibinfo  {journal} {Phys. Rev. Lett.}\ }\textbf {\bibinfo {volume} {96}},\
  \bibinfo {pages} {113002} (\bibinfo {year} {2006})}\BibitemShut {NoStop}%
\bibitem [{\citenamefont {Cuartero-Gonz\'alez}\ and\ \citenamefont
  {Fern\'andez-Dom\'inguez}(2018)}]{CuarFern18}%
  \BibitemOpen
  \bibfield  {author} {\bibinfo {author} {\bibfnamefont {A.}~\bibnamefont
  {Cuartero-Gonz\'alez}}\ and\ \bibinfo {author} {\bibfnamefont {A.~I.}\
  \bibnamefont {Fern\'andez-Dom\'inguez}},\ }\href {\doibase
  10.1021/acsphotonics.8b00678} {\bibfield  {journal} {\bibinfo  {journal} {ACS
  Photonics}\ }\textbf {\bibinfo {volume} {5}},\ \bibinfo {pages} {3415}
  (\bibinfo {year} {2018})},\ \Eprint
  {http://arxiv.org/abs/https://doi.org/10.1021/acsphotonics.8b00678}
  {https://doi.org/10.1021/acsphotonics.8b00678} \BibitemShut {NoStop}%
\bibitem [{\citenamefont {Cuartero-Gonz\'alez}\ and\ \citenamefont
  {Fern\'andez-Dom\'inguez}(2019)}]{CuarFern19}%
  \BibitemOpen
  \bibfield  {author} {\bibinfo {author} {\bibfnamefont {A.}~\bibnamefont
  {Cuartero-Gonz\'alez}}\ and\ \bibinfo {author} {\bibfnamefont {A.~I.}\
  \bibnamefont {Fern\'andez-Dom\'inguez}},\ }\href
  {https://arxiv.org/abs/1905.09893} {\bibfield  {journal} {\bibinfo  {journal}
  {arxiv}\ } (\bibinfo {year} {2019})}\BibitemShut {NoStop}%
\bibitem [{\citenamefont {Haus}(1984)}]{Haus}%
  \BibitemOpen
  \bibfield  {author} {\bibinfo {author} {\bibfnamefont {H.}~\bibnamefont
  {Haus}},\ }\href@noop {} {\emph {\bibinfo {title} {Waves and Fields in
  Optoelectronics}}}\ (\bibinfo  {publisher} {Prentice Hall},\ \bibinfo {year}
  {1984})\BibitemShut {NoStop}%
\bibitem [{\citenamefont {Fan}\ \emph {et~al.}(2003)\citenamefont {Fan},
  \citenamefont {Suh},\ and\ \citenamefont {Joannopoulos}}]{Fan2003}%
  \BibitemOpen
  \bibfield  {author} {\bibinfo {author} {\bibfnamefont {S.}~\bibnamefont
  {Fan}}, \bibinfo {author} {\bibfnamefont {W.}~\bibnamefont {Suh}}, \ and\
  \bibinfo {author} {\bibfnamefont {J.~D.}\ \bibnamefont {Joannopoulos}},\
  }\href@noop {} {\bibfield  {journal} {\bibinfo  {journal} {J. Opt. Soc. Am.
  B}\ }\textbf {\bibinfo {volume} {20}},\ \bibinfo {pages} {569} (\bibinfo
  {year} {2003})}\BibitemShut {NoStop}%
\bibitem [{\citenamefont {Suh}\ \emph {et~al.}(2004)\citenamefont {Suh},
  \citenamefont {Wang},\ and\ \citenamefont {Fan}}]{suh2004temporal}%
  \BibitemOpen
  \bibfield  {author} {\bibinfo {author} {\bibfnamefont {W.}~\bibnamefont
  {Suh}}, \bibinfo {author} {\bibfnamefont {Z.}~\bibnamefont {Wang}}, \ and\
  \bibinfo {author} {\bibfnamefont {S.}~\bibnamefont {Fan}},\ }\href@noop {}
  {\bibfield  {journal} {\bibinfo  {journal} {IEEE J. Quant. Electron.}\
  }\textbf {\bibinfo {volume} {40}},\ \bibinfo {pages} {1511} (\bibinfo {year}
  {2004})}\BibitemShut {NoStop}%
\bibitem [{\citenamefont {Leistikow}\ \emph {et~al.}(2009)\citenamefont
  {Leistikow}, \citenamefont {Johansen}, \citenamefont {Kettelarij},
  \citenamefont {Lodahl},\ and\ \citenamefont {Vos}}]{leistikow2009size}%
  \BibitemOpen
  \bibfield  {author} {\bibinfo {author} {\bibfnamefont {M.}~\bibnamefont
  {Leistikow}}, \bibinfo {author} {\bibfnamefont {J.}~\bibnamefont {Johansen}},
  \bibinfo {author} {\bibfnamefont {A.}~\bibnamefont {Kettelarij}}, \bibinfo
  {author} {\bibfnamefont {P.}~\bibnamefont {Lodahl}}, \ and\ \bibinfo {author}
  {\bibfnamefont {W.~L.}\ \bibnamefont {Vos}},\ }\href@noop {} {\bibfield
  {journal} {\bibinfo  {journal} {Phys. Rev. B}\ }\textbf {\bibinfo {volume}
  {79}},\ \bibinfo {pages} {045301} (\bibinfo {year} {2009})}\BibitemShut
  {NoStop}%
\bibitem [{\citenamefont {Yu}\ \emph {et~al.}(2003{\natexlab{b}})\citenamefont
  {Yu}, \citenamefont {Qu}, \citenamefont {Guo},\ and\ \citenamefont
  {Peng}}]{yu2003experimental}%
  \BibitemOpen
  \bibfield  {author} {\bibinfo {author} {\bibfnamefont {W.~W.}\ \bibnamefont
  {Yu}}, \bibinfo {author} {\bibfnamefont {L.}~\bibnamefont {Qu}}, \bibinfo
  {author} {\bibfnamefont {W.}~\bibnamefont {Guo}}, \ and\ \bibinfo {author}
  {\bibfnamefont {X.}~\bibnamefont {Peng}},\ }\href@noop {} {\bibfield
  {journal} {\bibinfo  {journal} {Chemistry of Materials}\ }\textbf {\bibinfo
  {volume} {15}},\ \bibinfo {pages} {2854} (\bibinfo {year}
  {2003}{\natexlab{b}})}\BibitemShut {NoStop}%
\bibitem [{\citenamefont {S\'{a}ez-Bl\'{a}zquez}\ \emph
  {et~al.}(2017)\citenamefont {S\'{a}ez-Bl\'{a}zquez}, \citenamefont {Feist},
  \citenamefont {Fern\'{a}ndez-Dom\'{i}nguez},\ and\ \citenamefont
  {Garc\'{i}a-Vidal}}]{SaezGarc17}%
  \BibitemOpen
  \bibfield  {author} {\bibinfo {author} {\bibfnamefont {R.}~\bibnamefont
  {S\'{a}ez-Bl\'{a}zquez}}, \bibinfo {author} {\bibfnamefont {J.}~\bibnamefont
  {Feist}}, \bibinfo {author} {\bibfnamefont {A.~I.}\ \bibnamefont
  {Fern\'{a}ndez-Dom\'{i}nguez}}, \ and\ \bibinfo {author} {\bibfnamefont
  {F.~J.}\ \bibnamefont {Garc\'{i}a-Vidal}},\ }\href {\doibase
  10.1364/OPTICA.4.001363} {\bibfield  {journal} {\bibinfo  {journal} {Optica}\
  }\textbf {\bibinfo {volume} {4}},\ \bibinfo {pages} {1363} (\bibinfo {year}
  {2017})}\BibitemShut {NoStop}%
\bibitem [{\citenamefont {Leatherdale}\ \emph
  {et~al.}(2002{\natexlab{b}})\citenamefont {Leatherdale}, \citenamefont {Woo},
  \citenamefont {Mikulec},\ and\ \citenamefont {Bawendi}}]{LeatBawe02}%
  \BibitemOpen
  \bibfield  {author} {\bibinfo {author} {\bibfnamefont {C.~A.}\ \bibnamefont
  {Leatherdale}}, \bibinfo {author} {\bibfnamefont {W.~K.}\ \bibnamefont
  {Woo}}, \bibinfo {author} {\bibfnamefont {F.~V.}\ \bibnamefont {Mikulec}}, \
  and\ \bibinfo {author} {\bibfnamefont {M.~G.}\ \bibnamefont {Bawendi}},\
  }\href {\doibase 10.1021/jp025698c} {\bibfield  {journal} {\bibinfo
  {journal} {J. Phys. Chem. B}\ }\textbf {\bibinfo {volume} {106}},\ \bibinfo
  {pages} {7619} (\bibinfo {year} {2002}{\natexlab{b}})}\BibitemShut {NoStop}%
\bibitem [{\citenamefont {Li}\ \emph {et~al.}(2014)\citenamefont {Li},
  \citenamefont {Chernikov}, \citenamefont {Zhang}, \citenamefont {Rigosi},
  \citenamefont {Hill}, \citenamefont {van~der Zande}, \citenamefont {Chenet},
  \citenamefont {Shih}, \citenamefont {Hone},\ and\ \citenamefont
  {Heinz}}]{LiHein14}%
  \BibitemOpen
  \bibfield  {author} {\bibinfo {author} {\bibfnamefont {Y.}~\bibnamefont
  {Li}}, \bibinfo {author} {\bibfnamefont {A.}~\bibnamefont {Chernikov}},
  \bibinfo {author} {\bibfnamefont {X.}~\bibnamefont {Zhang}}, \bibinfo
  {author} {\bibfnamefont {A.}~\bibnamefont {Rigosi}}, \bibinfo {author}
  {\bibfnamefont {H.~M.}\ \bibnamefont {Hill}}, \bibinfo {author}
  {\bibfnamefont {A.~M.}\ \bibnamefont {van~der Zande}}, \bibinfo {author}
  {\bibfnamefont {D.~A.}\ \bibnamefont {Chenet}}, \bibinfo {author}
  {\bibfnamefont {E.-M.}\ \bibnamefont {Shih}}, \bibinfo {author}
  {\bibfnamefont {J.}~\bibnamefont {Hone}}, \ and\ \bibinfo {author}
  {\bibfnamefont {T.~F.}\ \bibnamefont {Heinz}},\ }\href {\doibase
  10.1103/PhysRevB.90.205422} {\bibfield  {journal} {\bibinfo  {journal} {Phys.
  Rev. B}\ }\textbf {\bibinfo {volume} {90}},\ \bibinfo {pages} {205422}
  (\bibinfo {year} {2014})}\BibitemShut {NoStop}%
\bibitem [{\citenamefont {Hakami}\ \emph {et~al.}(2014)\citenamefont {Hakami},
  \citenamefont {Wang},\ and\ \citenamefont {Zubairy}}]{HakaZuba14}%
  \BibitemOpen
  \bibfield  {author} {\bibinfo {author} {\bibfnamefont {J.}~\bibnamefont
  {Hakami}}, \bibinfo {author} {\bibfnamefont {L.}~\bibnamefont {Wang}}, \ and\
  \bibinfo {author} {\bibfnamefont {M.~S.}\ \bibnamefont {Zubairy}},\ }\href
  {\doibase 10.1103/PhysRevA.89.053835} {\bibfield  {journal} {\bibinfo
  {journal} {Phys. Rev. A}\ }\textbf {\bibinfo {volume} {89}},\ \bibinfo
  {pages} {053835} (\bibinfo {year} {2014})}\BibitemShut {NoStop}%
\bibitem [{\citenamefont {Dzsotjan}\ \emph {et~al.}(2016)\citenamefont
  {Dzsotjan}, \citenamefont {Rousseaux}, \citenamefont {Jauslin}, \citenamefont
  {des Francs}, \citenamefont {Couteau},\ and\ \citenamefont
  {Gu\'erin}}]{DzsoGuer16}%
  \BibitemOpen
  \bibfield  {author} {\bibinfo {author} {\bibfnamefont {D.}~\bibnamefont
  {Dzsotjan}}, \bibinfo {author} {\bibfnamefont {B.}~\bibnamefont {Rousseaux}},
  \bibinfo {author} {\bibfnamefont {H.~R.}\ \bibnamefont {Jauslin}}, \bibinfo
  {author} {\bibfnamefont {G.~C.}\ \bibnamefont {des Francs}}, \bibinfo
  {author} {\bibfnamefont {C.}~\bibnamefont {Couteau}}, \ and\ \bibinfo
  {author} {\bibfnamefont {S.}~\bibnamefont {Gu\'erin}},\ }\href {\doibase
  10.1103/PhysRevA.94.023818} {\bibfield  {journal} {\bibinfo  {journal} {Phys.
  Rev. A}\ }\textbf {\bibinfo {volume} {94}},\ \bibinfo {pages} {023818}
  (\bibinfo {year} {2016})}\BibitemShut {NoStop}%
\bibitem [{\citenamefont {Stockman}(2011)}]{Stockman11}%
  \BibitemOpen
  \bibfield  {author} {\bibinfo {author} {\bibfnamefont {M.~I.}\ \bibnamefont
  {Stockman}},\ }\href {\doibase 10.1364/OE.19.022029} {\bibfield  {journal}
  {\bibinfo  {journal} {Opt. Express}\ }\textbf {\bibinfo {volume} {19}},\
  \bibinfo {pages} {22029} (\bibinfo {year} {2011})}\BibitemShut {NoStop}%
\end{thebibliography}%

\end{document}